\documentclass[10pt,preprint]{aastex}
\begin{document}

\slugcomment{Accepted to the {\it Astrophysical Journal}}

\shorttitle{On the MBM12 Young Association}
\shortauthors{Luhman}

\title{On the MBM12 Young Association}

\author{K. L. Luhman}
\affil{Harvard-Smithsonian Center for Astrophysics, 60 Garden St., 
Cambridge, MA 02138}

\email{kluhman@cfa.harvard.edu}

\begin{abstract}

I present a comprehensive study of the MBM12 young association (MBM12A). 
By combining infrared (IR) photometry from the Two-Micron All-Sky Survey 
(2MASS) survey with new optical imaging and spectroscopy, I have performed a 
census of the MBM12A membership that is complete to 0.03~$M_{\odot}$ ($H\sim15$)
for a $1.75\arcdeg\times1.4\arcdeg$ field encompassing the MBM12 cloud. I find 
five new members with masses of 0.1-0.4~$M_{\odot}$ and a few
additional candidates that have not been observed spectroscopically.
From an analysis of optical and IR photometry for stars in the direction of 
MBM12, I identify M dwarfs in the foreground and background of the cloud.
By comparing the magnitudes of these stars to those of local field dwarfs, I
arrive at a distance modulus $7.2\pm0.5$ (275~pc) to the MBM12 cloud; 
it is not the nearest molecular cloud and is not inside the local bubble of hot 
ionized gas as had been implied by previous distance estimates of 50-100~pc.
I have also used Li strengths and H-R diagrams to constrain the absolute and
relative ages of MBM12A and other young populations; these data indicate
ages of $2^{+3}_{-1}$~Myr for MBM12A and $\sim10$~Myr for the TW~Hya and 
$\eta$~Cha associations. 
MBM12A may be a slightly evolved version of the aggregates of young stars 
within the Taurus dark clouds ($\sim1$~Myr) near the age of the IC~348 
cluster ($\sim2$~Myr).

\end{abstract}


\section{Introduction}

Young nearby stellar populations offer unique advantages in the study of 
star and planet formation.
Because of the youth and proximity of the members of these regions,
substellar companions and circumstellar material are bright and can be 
observed at small separations and high spatial resolution. 
In particular, associations at ages of 1 to 10~Myr within 100~pc should contain
the brightest brown dwarfs in the sky and giant planets that are emerging 
from their natal disks.

The nearest star forming regions and young open clusters are 
at distances of $\sim150$~pc, which have been identified by their associated 
molecular clouds, infrared (IR) excess emission, and compact and rich
populations. At closer distances, associations with small memberships and 
ages of $\gtrsim5$~Myr are less conspicuous and can go unnoticed. 
Recently, data from the {\it Hipparcos} and
{\it ROSAT} missions have been used in searching for members of previously
undiscovered nearby associations. These observations have revealed four 
populations at ages of 10-30~Myr and distances of 50-100~pc, namely the 
TW~Hya (TWA) (Kastner et al.\ 1997; Webb et al.\ 1999; Zuckerman et al.\ 2001), 
$\eta$~Chamaeleonits (Mamajek et al.\ 1999), Horologium (HorA) 
(Torres et al.\ 2000), 
and Tucana associations (Zuckerman \& Webb 2000). The group of T~Tauri stars 
projected against cloud~12 from Magnani, Blitz, \& Mundy (1985) (MBM12)
has been suggested as an additional young association within 100~pc, and 
is the subject of this study.

From distance estimates of 60-70~pc (Hobbs, Blitz, \& Magnani 1986) and 
53-102~pc (Hearty et al.\ 2000a), the MBM12 cloud, also known as L1453, 
L1454, L1457, and L1458 from Lynds (1962), is believed to be the nearest 
molecular cloud. 
The line-of-sight extinctions through the cloud are typically $A_V=0$-5 and
can reach as high as $A_V\sim10$ in a few locations (Zuckerman et al.\ 1992;
Hearty et al.\ 2000b, hereafter H00b), placing it near the boundary between
translucent and dark clouds ($A_V=5$, van Dishoeck \& Black 1988).
Surveys in H$\alpha$ and X-rays have identified seven young stars 
near the MBM12 cloud
(Herbig \& Rao 1972; Stephenson 1986; Downes \& Keyes 1988; Herbig \& Bell 
1988; Fleming, Gioia, \& Maccacaro 1989; Mart{\'\i}n et al.\ 1994; 
Fern\'{a}ndez et al.\ 1995; Magnani et al.\ 1995; H00b).
An additional X-ray source in the vicinity of the MBM12 cloud, RXJ0306.5+1921, 
was proposed as young member of MBM12A by H00b, but it is probably an older 
object that is not related to these T~Tauri stars, as shown in this work.
Based on the strengths of Li absorption and H$\alpha$ and mid-IR emission 
and the presence of the parent molecular cloud, the group of stars 
towards MBM12 is believed to be younger than TWA and $\eta$~Cha ($\sim10$~Myr) 
with an age of a few million years (H00b; Jayawardhana et al.\ 2001).

In this paper, I present the results of a magnitude-limited search for new
members of the MBM12 association (MBM12A) that is complete to 0.03~$M_{\odot}$.
Following the methods of similar searches in Taurus by Brice\~{n}o
et al.\ (1998) and Luhman (2000), I combine $R$ and $I$ images of the
dark cloud with IR photometry from the Two-Micron All-Sky Survey (2MASS) and
select candidate members for spectroscopy, from which five new members are
found. In addition, I revise the distance to MBM12A, use Li data and the 
Hertzsprung-Russell (H-R) diagram to estimate the absolute and relative ages
of this and other young associations, and discuss the disk properties of stars
in MBM12A.

\section{Observations and Data Analysis}
\label{sec:obs}

\subsection{Photometry}
\label{sec:obs1}

Images of the MBM12 cloud were obtained with the four shooter camera at the
Fred Lawrence Whipple Observatory 1.2~m telescope on 2000 October 24-26.
The instrument contained four $2048\times2048$
CCDs separated by $\sim45\arcsec$ and arranged in a $2\times2$ grid. After
binning $2\times2$ during readout, the plate scale was $0\farcs67$~pixel$^{-1}$.
The survey area for MBM12 was divided into a $5\times4$ grid in right 
ascension and declination where the grid cells were separated by $20\arcmin$. 
At each grid cell, two positions separated by $40\arcsec$ in right 
ascension and declination were observed. 
At each position, images were obtained with exposure
times of 15 and 450~s at $R$ and 10 and 300~s at $I$.  
The images were bias subtracted, divided by dome flats, registered, and 
combined into one image at each band and exposure time. 
Stars in overlapping regions of the frames were used to scale all images to
the frames that were known to be photometric. Photometry and image coordinates
of the sources in these data were measured with DAOFIND and 
PHOT under the IRAF package APPHOT.
For most stars, aperture photometry was extracted with a radius of four pixels.
Smaller apertures were used for closely spaced stars. The background
level was measured in an annulus around each source and subtracted from
the photometry, where the inner radius of the annulus was five pixels and the
width was one pixel.
The photometry was calibrated in the Cousins $I$ system through observations
of standards across a range of colors (Landolt 1992). 
Saturation in the short exposures occurred at $R=11.5$-12 and $I=11$-11.5. 
The completeness limits of the long exposures were $R\sim20.5$ and $I\sim19.25$.
Typical photometric uncertainties (1~$\sigma$) are 0.04~mag at $R=20$
and $I=18.75$ and 
0.1~mag at $R=21$ and $I=19.75$. The plate solution was derived from
coordinates of sources observed in the 2MASS Spring 1999 Release Point Source 
Catalog that appeared in the optical images and were not saturated. 

\begin{figure}
\plotone{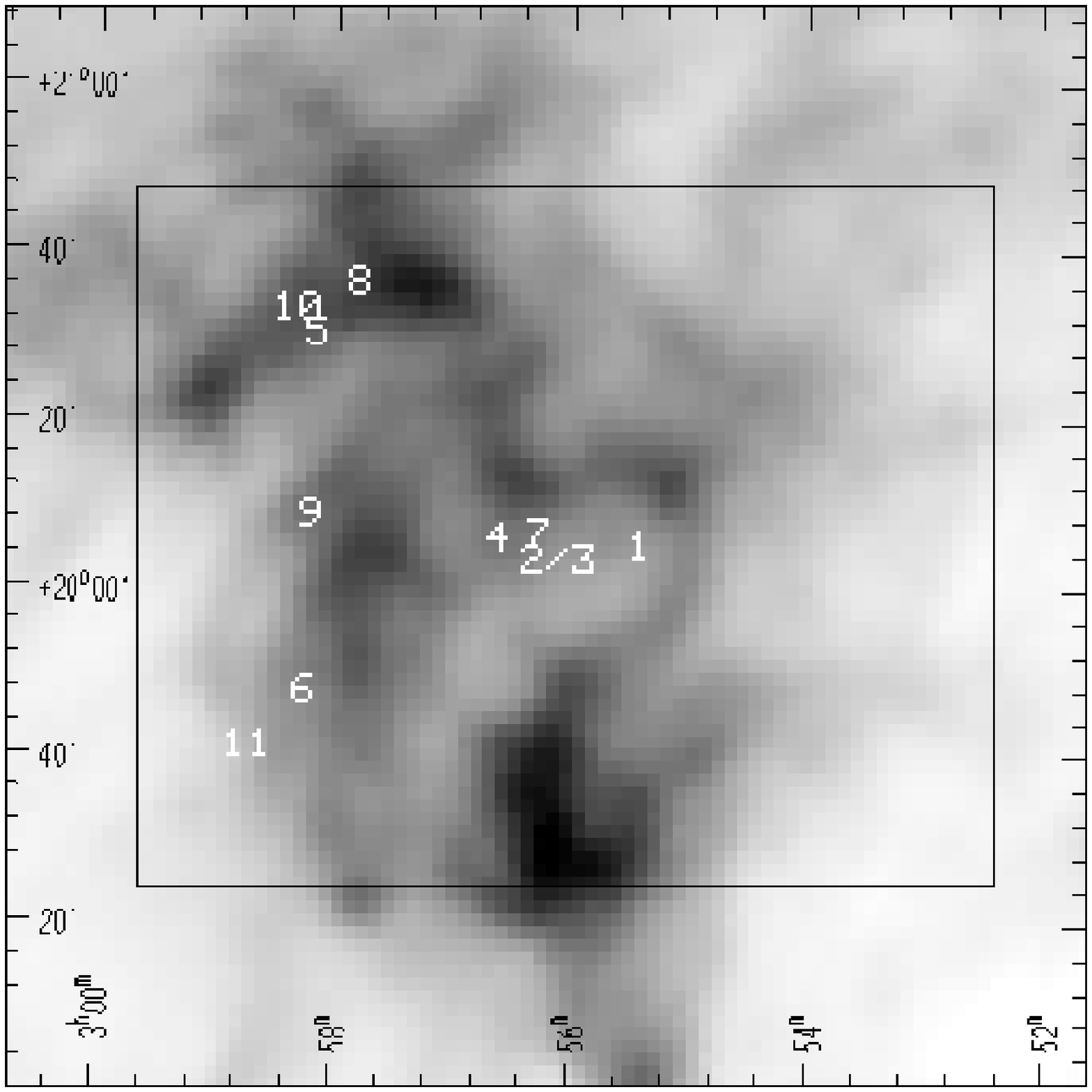}
\caption{
The {\it IRAS} 100~$\mu$m image of the MBM12 cloud is shown with the 
positions of previously known young stars (MBM12A~1-6) and the new ones 
discovered in this work (MBM12A~7-11). 
MBM12A~2 and 3 (LkH$\alpha$~262 and 263) are a wide binary system. 
A young star outside of the region shown here, MBM12A~12 (S18), is 
located $2.5\arcdeg$ southeast of the MBM12 cloud and is a likely member 
of this association.
Images at $R$ and $I$ were obtained for the $1.75\arcdeg\times1.4\arcdeg$ area 
within the rectangle, which is defined as 
$\alpha=2^{\rm h}52^{\rm m}26\fs4$ to $2^{\rm h}59^{\rm m}40^{\rm s}$, 
$\delta=19\arcdeg24\arcmin24\arcsec$ to $20\arcdeg47\arcmin37\arcsec$ (J2000).
}
\label{fig:map1}
\end{figure}

The boundaries of the rectangular field from which photometry was extracted are 
$\alpha=2^{\rm h}52^{\rm m}26\fs4$, $2^{\rm h}59^{\rm m}40^{\rm s}$ and
$\delta=19\arcdeg24\arcmin24\arcsec$, $20\arcdeg47\arcmin37\arcsec$ (J2000).
This region is indicated in Figure~\ref{fig:map1}.  Out of 
4661 2MASS sources within this field, 50 objects were not measured
in the optical data because they fell within gaps between the detectors or
were contaminated by bad pixels; these 2MASS sources are excluded from the 
study. In addition, 5 known asteroids that were flagged in the 2MASS database 
were rejected, leaving a total of 4606 2MASS sources. 
Out of the 9157 optical sources, 4471 stars have 2MASS detections, resulting
in a total of 9292 optical and IR sources. 
When available, the final coordinates are those measured from the
optical data. For the other stars -- the 44 2MASS sources that were saturated 
or the 91 2MASS sources that were too red to be detected in the optical 
images -- the 2MASS coordinates are adopted.  Photometry and coordinates are
given in Tables~\ref{tab:mem}-\ref{tab:cand} for stars spectroscopically 
identified as young members, foreground dwarfs, or background stars and 
for candidate members without spectra. The complete list of optical and
IR photometry for the MBM12 cloud can be obtained from the author upon request. 

\subsection{Spectroscopy}
\label{sec:obs2}

In \S~\ref{sec:ident}, the optical and IR photometry for the MBM12 cloud is used
to identify candidate members of the young association. The spectroscopic 
observations of those candidates is now described. For the remainder of this 
paper, the previously known members and the new members discovered in 
this work are named by the MBM12A designations defined in Table~\ref{tab:mem}.

Candidates of intermediate brightness ($I=13$-17) were observed with 
moderate-resolution spectroscopy at the Steward Observatory 2.3~m Bok 
Reflector. On 2000 December 4, spectra were obtained for 22 candidate members, 
which include the new members MBM12A~7-11, 11 of the foreground 
dwarfs in Table~\ref{tab:fore}, and 6 of the background stars in 
Table~\ref{tab:back}. 
The previously known members MBM12A~3 and 6 were observed as well.
The Boller and Chivens Spectrometer was operated with the 400~l~mm$^{-1}$ 
grating ($\lambda_{\rm blaze}=7506$~\AA) and Y48 blocking filter 
to obtain spectra from 5900 to 9300~\AA. 
The spectra were obtained with the slit rotated to the parallactic angle for 
each target. 
The slit width was $1\farcs5$, producing a spectral resolution of FWHM=7.0~\AA. 
The exposure times ranged from 300 to 1800~s.

Out of the 16 candidates from December~4 that showed K or M-type spectra,
the 12 brightest ($I=13$-15) were observed at higher resolution near 
the Li~6707~\AA\ transition on 2000 December 5.  These sources consist of the
new members MBM12A~7 and 10 and 10 of the field dwarfs in 
Table~\ref{tab:fore}.
In addition, 34 of the brightest remaining candidates ($I=11$-13) --
4 field dwarfs and 30 background stars -- and 
the previously known members MBM12A~3 and 4 were observed. 
The 832~l~mm$^{-1}$ grating ($\lambda_{\rm blaze}=7524$~\AA) and Y48 blocking
filter provided a 
wavelength coverage of 6100-7800~\AA. With the $1\farcs5$ slit, the spectral
resolution was 3.1~\AA. The exposure times ranged from 120 to 1800~s.

The faintest three candidates ($I=17$-19) were observed with low-resolution 
spectroscopy at the Keck~I telescope. Spectra were obtained for 
2MASSs~J0253030+194904 on 2000 November 25 and for 2MASSs~J0258110+192908 
and 2MASSs~J0257299+204308 on 2000 November 26 with the Keck low-resolution 
imaging spectrometer (LRIS; Oke et al.\ 1995).  
The long-slit mode of LRIS was 
used with the 150~l~mm$^{-1}$ grating ($\lambda_{\rm blaze}=7500$~\AA) and
GG570 blocking filter. The maximum wavelength coverage of LRIS, 3800 
to 11000~\AA, was provided in one grating setting centered near 7500~\AA. 
The spectra were obtained with the slit rotated to the parallactic angle
for each target. The slit widths were $0\farcs7$ for 2MASSs~J0253030+194904 and 
$1\farcs0$ for 2MASSs~J0258110+192908 and 2MASSs~J0257299+204308, 
producing spectral resolutions of 16 and 20~\AA, respectively.
The exposure times were 600~s for 2MASSs~J0253030+194904 and
1200~s for 2MASSs~J0258110+192908 and 2MASSs~J0257299+204308.

Standard data reduction procedures were followed for all spectra.
After bias subtraction and flat-fielding
with internal continuum lamps, the spectra were extracted and calibrated in
wavelength with arc lamp data. The spectra were then corrected 
for the sensitivity functions measured from observations of the 
spectrophotometric standard stars Feige~11 and 34. 

\section{New Members of MBM12A}

\subsection{Identification of Candidate Members}
\label{sec:ident}

Candidate members of MBM12A can be identified through optical and IR 
color-color and color-magnitude diagrams.
In an optical color-magnitude diagram, the members of a nearby young population 
form a well-defined locus that is above most of the background stars. 
In a similar search for young low-mass stars, 
Luhman (2000) supplemented the $R$ and $I$ photometry of Brice\~{n}o et 
al.\ (1998) with $I$ and $z\arcmin$ data to provide completeness to 
substellar masses for the full range of the typical reddenings of Taurus 
members ($A_V=0$-10). For the MBM12 cloud, the total extinction 
is relatively modest ($A_V<5$) at most sight lines and the 
reddenings of the previously known members are even lower ($A_V<2$).
As a result, the $R$ and $I$ bands are sufficient for this search for new 
members of MBM12A. 

\begin{figure}
\plotone{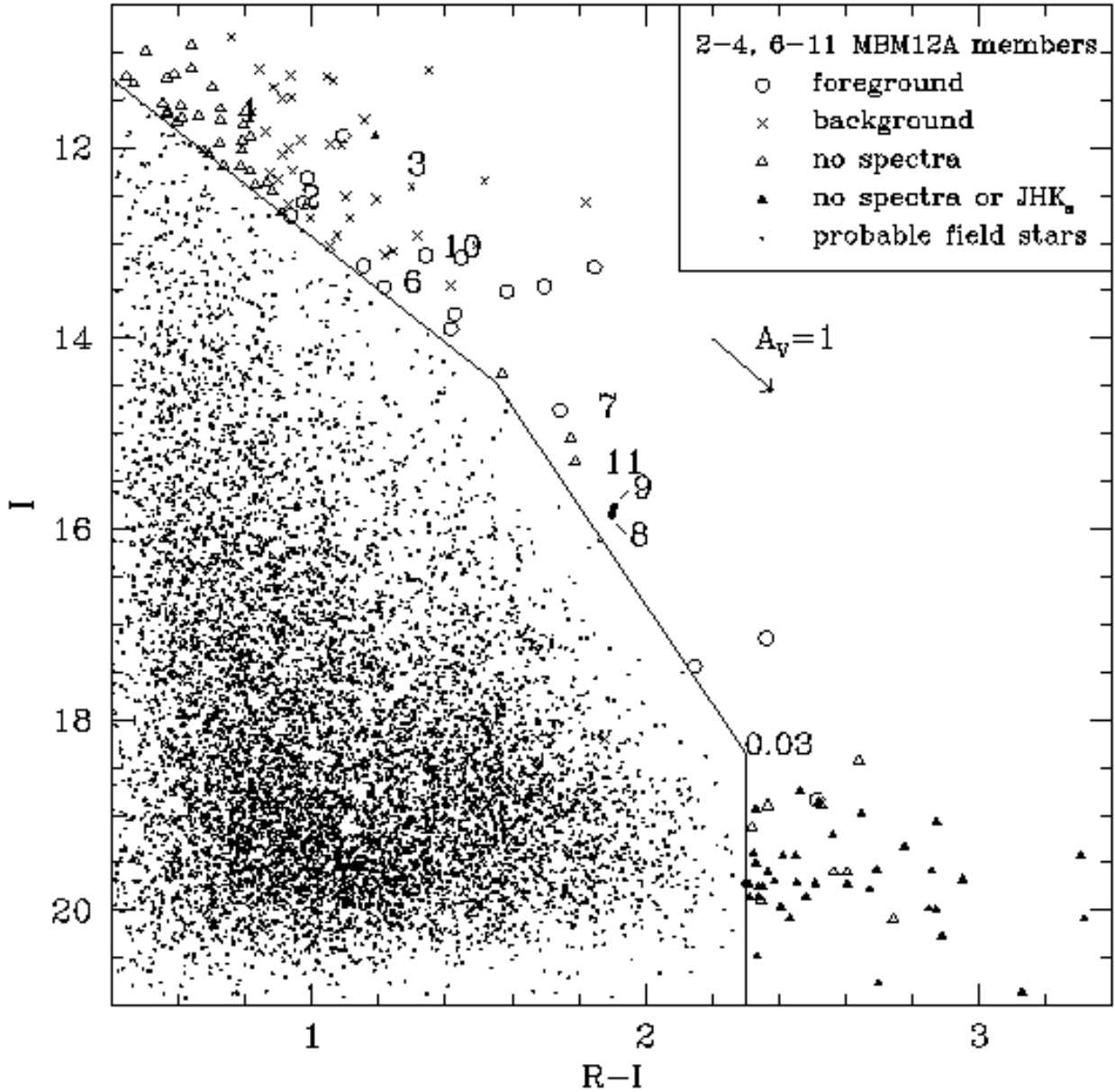}
\caption{
$R-I$ vs.\ $I$ for the MBM12 cloud.
MBM12A~2, 3, 4, and 6 are previously known young stars 
(MBM12A~1 and 5 are saturated at $R$).
Stars above the solid boundary are candidate members of this young
association while sources below the boundary are probable field stars.
Spectroscopy of the candidates has been used to identify new young 
members (MBM12A~7-11), foreground dwarfs ({\it circles}), and background stars 
({\it crosses}). The remaining 
stars above the boundary that lack spectra are divided into those with and 
without IR photometry ({\it open and solid triangles}).
The unreddened position of 0.03~$M_{\odot}$ ($\sim$M8) is shown for a distance
modulus of 7.2 (275~pc) and an age of 2~Myr (Baraffe et al.\ 1998). 
}
\label{fig:ri}
\end{figure}

I now describe the region in Figure~\ref{fig:ri} where stellar and substellar
members of MBM12A are expected to reside.
An isochrone in $R-I$ vs.\ $I$ can be constructed by 
combining the models of Baraffe et al.\ (1998) with the appropriate temperature 
and color conversions and bolometric corrections (see \S~\ref{sec:mbm12}).
While the vertical offset of this isochrone depends on the assumed age 
and distance, the shape remains fairly constant for ages of 1-30~Myr. 
Therefore, independent of the precise age or distance, all members of MBM12A
should fall above a boundary that is parallel to a young model isochrone 
and is below the observed positions of previously known members. 
Because the colors of the latest M stars saturate near $R-I=2.4$-2.5,  
the boundary becomes vertical at $R-I=2.3$ in Figure~\ref{fig:ri}. 

\begin{figure}
\plotone{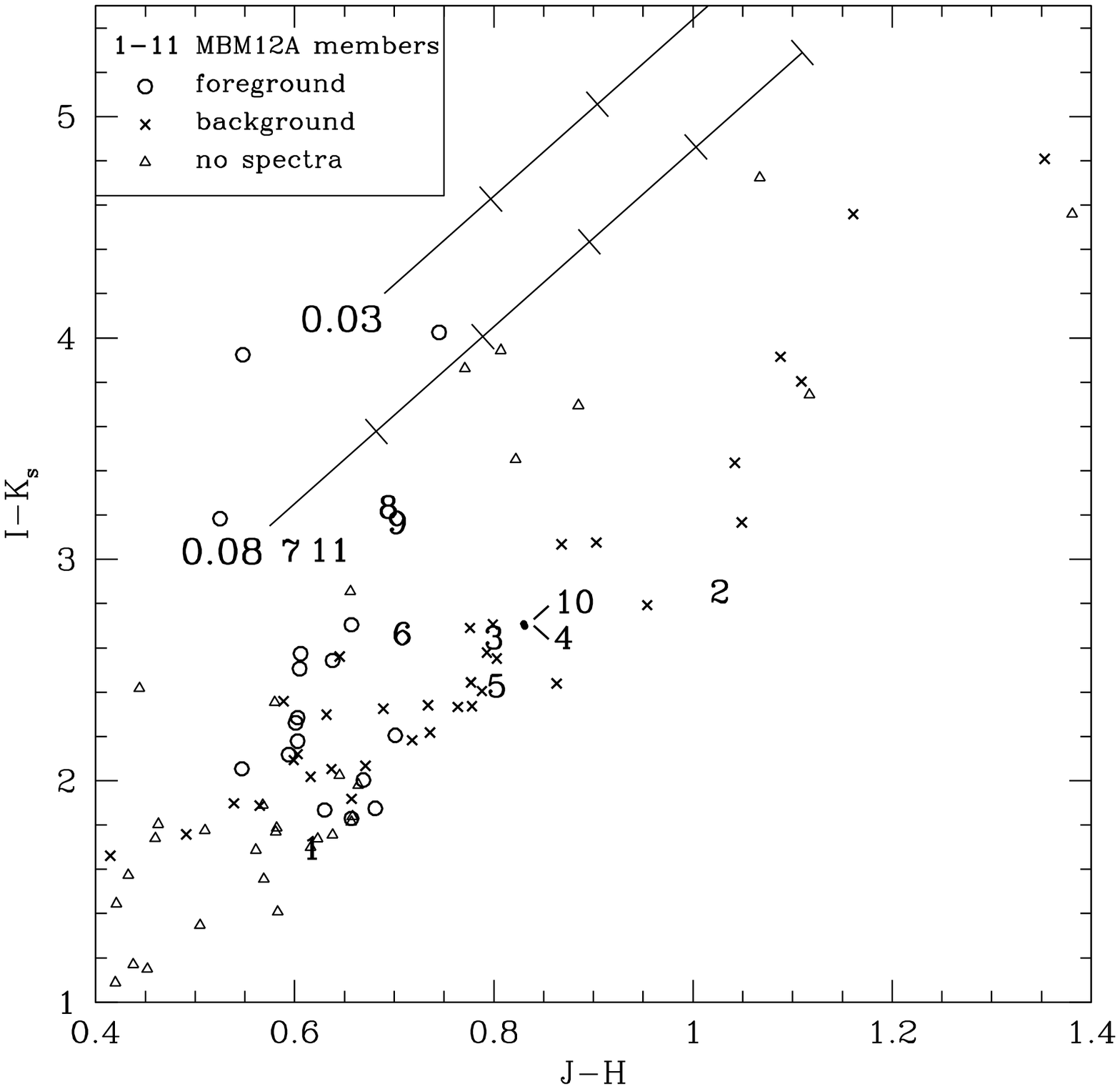}
\caption{
$J-H$ vs.\ $I-K_s$ for the MBM12 cloud.
MBM12A~1-6 are previously known young stars.
Spectroscopy of candidate members from $R-I$ vs.\ $I$ in Figure~\ref{fig:ri} 
has been used to identify new members (MBM12A~7-11), foreground dwarfs 
({\it circles}), and background stars ({\it crosses}).
The remaining stars that were not rejected as field stars
in Figure~\ref{fig:ri} and that lack spectra are shown ({\it open triangles}), 
while the probable field stars from that diagram are omitted. 
The lower and upper lines represent the reddening vectors 
for spectral types of M6.5 and M8 (Leggett 1992), which correspond to 0.08 and
0.03~$M_{\odot}$ for an age of 2~Myr (Baraffe et al.\ 1998). The reddening
vectors are marked at intervals of $A_V=1$.
}
\label{fig:jhik}
\end{figure}

Candidate members in a given region of the above color-magnitude diagram can be
reddened background stars or K and M-type stars, where the latter are either 
members of the association or foreground dwarfs. Many of the background
stars can be separated from the late-type stars and rejected from the 
candidate list through color-color diagrams. For instance, 
the diagram of $J-H$ vs.\ $I-K_s$ in Figure~\ref{fig:jhik} is useful for 
colors at $R-I\gtrsim2$.
With progressively later M spectral types, $I-K_s/J-H$ quickly increases
and the corresponding reddening vector separates from the band occupied by
most stars, which are reddened background stars of earlier spectral types. 
The combination of the models of Baraffe et al.\ (1998) and the compatible
temperature scale of Luhman (1999) indicate that brown dwarfs at masses of
0.03 and 0.08~$M_{\odot}$ and an age of 2~Myr should have spectral types near
M8 and M6.5, respectively.  
For the filters used here, Luhman (2000) derived the reddening vector 
$E(I-K_s)/E(J-H)=4.0$. 
By combining this slope with the dwarf colors of Leggett (1992) 
for spectral types of M8 and M6.5, I plot the reddening vectors for 0.03 and 
0.08~$M_{\odot}$ in Figure~\ref{fig:jhik}. Only three of the candidates with
$R-I>2$ from Figure~\ref{fig:ri} exhibit the $I-K_s$ and $J-H$ 
colors expected for late M stars.

Finally, as shown in \S~\ref{sec:dwarfs}, a diagram of $R-I$ vs.\ $I-K_s$ 
is a powerful tool for separating background stars and M-type stars at $R-I>1$. 
If this method had been used here in developing the candidate list, 
the spectroscopy of background stars at $R-I>1$ would have been avoided. 

\subsection{Spectral Classification of Candidate Members}
\label{sec:sptypes}

The candidate members of MBM12A that were observed spectroscopically are
indicated in Figure~\ref{fig:ri}.
As described in \S~\ref{sec:obs2}, the candidates with $I=13$-15
were observed first at low resolution. If they exhibited the K and M-type
spectra expected for members (e.g., strong TiO and VO absorption bands), 
they were observed at higher resolution to check for the strong Li absorption 
that would identify them as young members rather than foreground field dwarfs. 
The low-resolution spectra were unnecessary for the brightest candidates
as they could be efficiently observed at the higher resolution. 
At the spectral types of $>$M5 expected for the faintest and reddest candidates 
($I>15$, $R-I>1.8$), the K and Na absorption lines vary significantly
between dwarfs and pre-main-sequence stars (Mart{\'\i}n, Rebolo, \& 
Zapatero Osorio 1996; Luhman et al.\ 1998a, 1998b; Luhman 1999).
Because these features are easily detected in low-resolution spectra, 
higher-resolution data for Li were unnecessary for distinguishing young 
members from field dwarfs among these faintest candidates.

The targets in the spectroscopic sample can be background stars, foreground
stars, or young members of MBM12A. 
Background stars are predominantly
giants and early-type stars; in low-resolution spectra they exhibit only
a few absorption lines (H$\alpha$, Ca~II) and are otherwise featureless.
The 36 stars with these spectral characteristics of background objects are 
listed in Table~\ref{tab:back}. Background dwarfs should not 
appear in a spectroscopic sample of this type because they fall well below 
the locus of young members in $R-I$ vs.\ $I$. However, in the initial 
data reduction used for selecting spectroscopic candidates immediately following
the imaging observations, bad pixels near 2MASSs~J0257299+204308 caused the
star to appear as a candidate member. After correcting for the bad pixels in
the final data reduction, the star falls below the boundary in 
Figure~\ref{fig:ri}. The spectrum of this object is consistent with that of a 
background dwarf with a spectral type of M2.75V and a reddening of $A_V\sim4$-5.

Foreground dwarfs and young members of MBM12A can 
be distinguished from each other by the Li, K, and Na absorption features. 
However, the Li~I 6707.8~\AA\ transition must be used with caution as 
a youth diagnostic (Brice\~{n}o et al.\ 1997). The strength of the line
can be significantly overestimated in low-resolution data because of 
blending with nearby Fe lines (Basri, Mart{\'\i}n, \& Bertout 1991; 
Covino et al.\ 1997).  In addition, the transition is less useful as a youth 
indicator for stars earlier than $\sim$K0.
Comparisons of Li measurements of the same stars at both low 
and high spectral resolutions suggest that accurate equivalent widths 
can be obtained at a resolution of FWHM$\lesssim1.5$~\AA\ while the strength of
the line is overestimated for FWHM$\gtrsim3.5$~\AA\ (Walter et al.\ 1988;
G\'{o}mez et al.\ 1992; Brice\~{n}o et al.\ 1997; Covino et al.\ 1997; 
Favata et al.\ 1997; Neuh\"{a}user et al.\ 1997).
The errors are greatest for G and K-type stars while the contaminating 
lines are weak relative to Li for M-type stars. In this study of MBM12A,
given the resolution of 3.1~\AA\ for the Li spectra and the M types of most
of the non-background candidates, the resulting Li measurements should be 
fairly reliable. This conclusion can be tested by comparing the Li 
strengths measured here with the equivalent widths
found at higher resolution by H00b for MBM12A~4 and by 
Jayawardhana et al.\ (2001) for MBM12A~3 and 4. 
As shown in Table~\ref{tab:lines}, the
Li measurements in this study are roughly consistent with previous 
data at higher resolution and should overestimate the line strengths by 
no more than $\sim10$\%. 

\begin{figure}
\epsscale{0.75}
\plotone{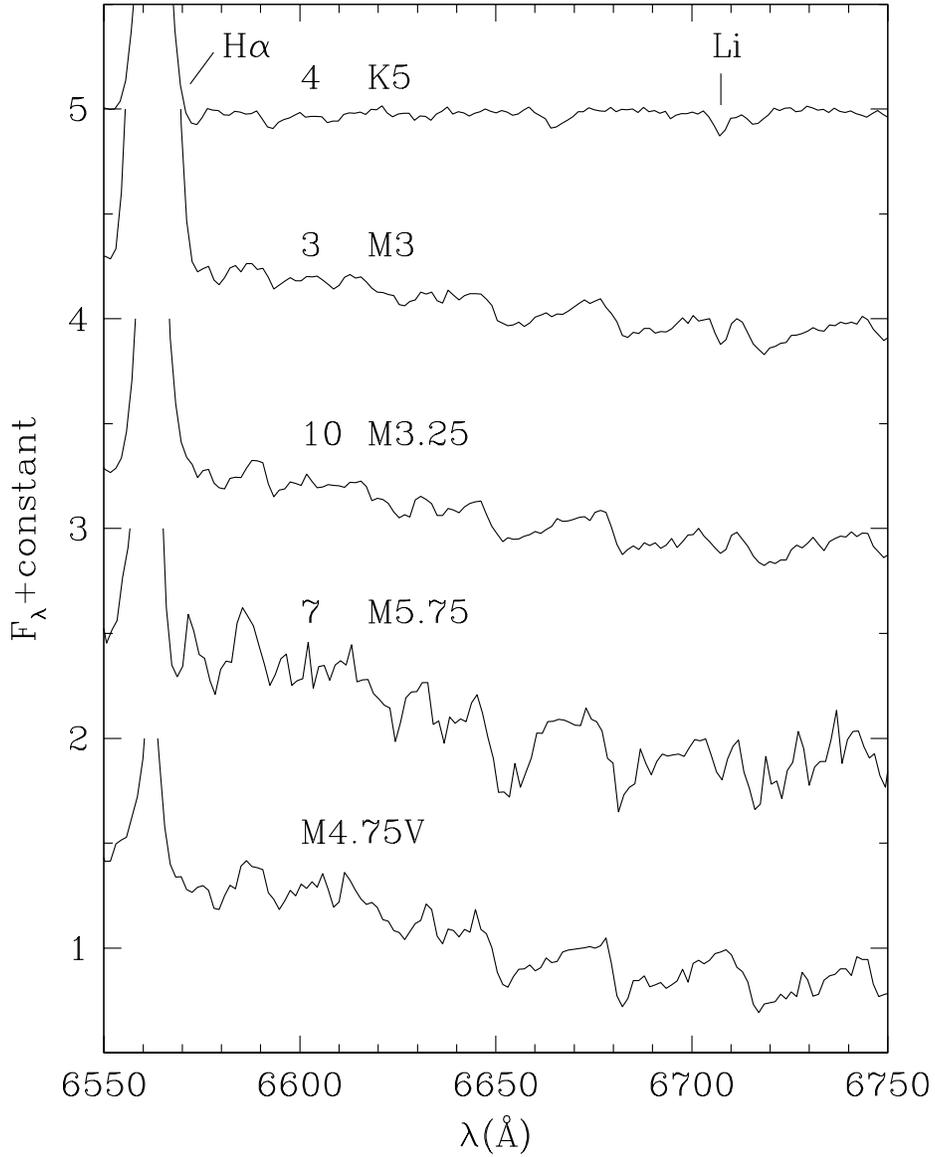}
\caption{
Medium-resolution spectra of the H$\alpha$ and Li 6707.8~\AA\ transitions
for stars towards the MBM12 cloud. 
MBM12A~3 and 4 are previously known young stars and MBM12A~7 and 10 are
new ones discovered in this work,
all of which exhibit the strong Li absorption that is 
expected for young late-type stars. 
Meanwhile, the lack of detectable Li absorption ($<0.1$~\AA) in the
bottom star (2MASSs~J0253031+192907) is indicative of a field dwarf. 
All data have a resolution of FWHM=3.1~\AA\ and are normalized to the 
continuum near the Li line.
}
\label{fig:specli}
\end{figure}

\begin{figure}
\epsscale{0.9}
\plotone{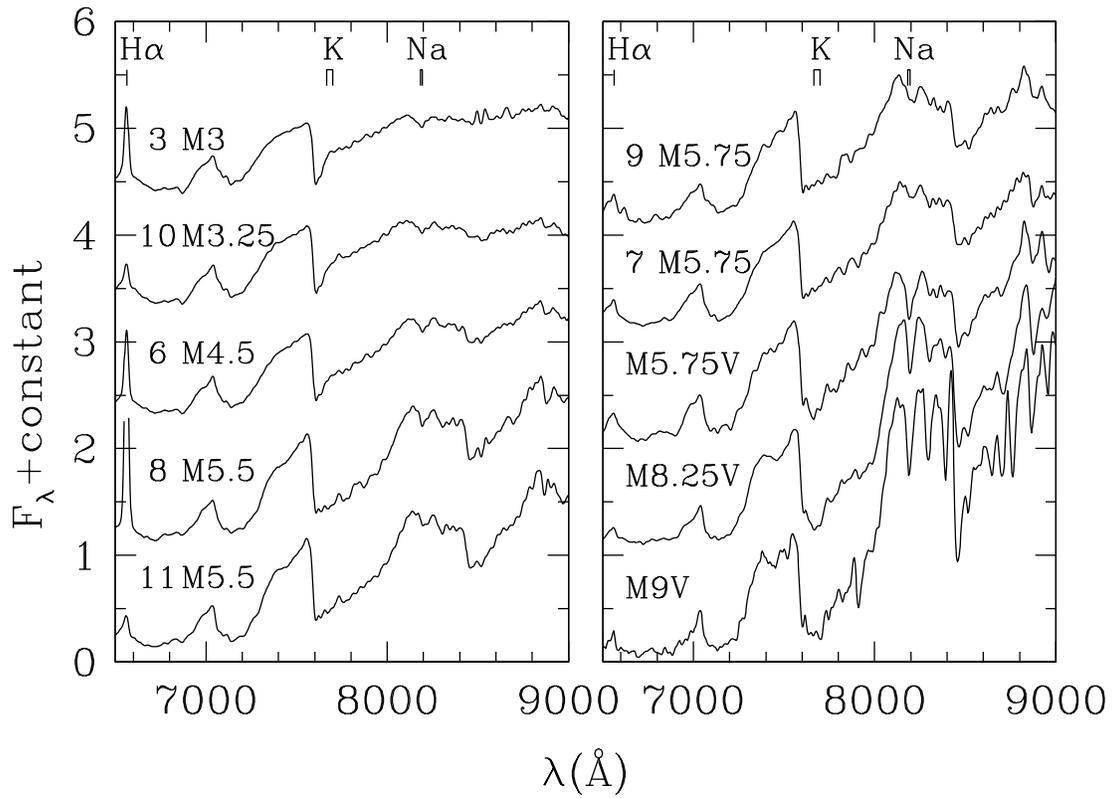}
\caption{
Low-resolution spectra of stars towards the MBM12 cloud. 
MBM12A~3 and 6 are previously known young stars and MBM12A~7-11 are the new
ones discovered in this work. 
The strong K and Na absorption in the spectra of the last three stars indicates 
that they are field dwarfs rather than young late-type members of the 
association. 
All data are smoothed to a resolution of 25~\AA\ and normalized at 7500~\AA.
}
\label{fig:spec}
\end{figure}

No Li absorption is detected for 14 of the 
candidates showing M-type spectra.  The upper limits for Li absorption were 
$W_{\lambda}<0.2$ for 2MASSs~J0257501+195830 and 2MASS~J0253243+200659 and
$W_{\lambda}<0.05$-0.1~\AA\ for the remaining stars. A sample spectrum near Li 
is shown for one of these dwarfs in Figure~\ref{fig:specli}. Strong absorption 
in K and Na is also found in the spectra of these stars. All of these spectral
properties are indicative of field M dwarfs. 
The lack of reddening in their spectra and colors and their position above
the main sequence at the distance to MBM12A suggest that these stars are in
the foreground of the cloud.
For the faint candidates 2MASSs~J0255083+193859, 2MASSs J0258110+192908, and
2MASSs J0253030+194904, only low-resolution spectra are available. However,
even without Li measurements, these three stars are clearly M dwarfs by 
the strong K and Na absorption in their spectra in Figure~\ref{fig:spec}.
Spectral types listed in Table~\ref{tab:fore} for these 17 foreground
stars were measured by comparison to the spectra of standard dwarfs from
Kirkpatrick, Henry, \& McCarthy (1991), Henry, Kirkpatrick, \& Simons (1994), 
and Kirkpatrick, Henry, \& Irwin (1997).

The strong Li absorption expected for young late-type stars 
($W_{\lambda}\gtrsim0.4$~\AA) is found in the spectra of the candidates
MBM12A~7 and 10, as well as the previously known members MBM12A~3 and 4,
as shown in Figure~\ref{fig:specli}. Measurements of Li and 
H$\alpha$ for these stars are given in Table~\ref{tab:lines}.
The spectra of MBM12A~7 and 10 also exhibit
weak K and Na lines relative to dwarfs of the same spectral types, which
is further evidence of their pre-main-sequence nature. The three
candidates MBM12A~8, 9, and 11 were observed with low-resolution 
spectroscopy but were too faint for higher resolution Li measurements. 
In Figure~\ref{fig:spec}, the spectra of these stars show the weak K and
Na absorption features that are signatures of pre-main-sequence stars. 
The moderately strong H$\alpha$ emission (10-15~\AA) in MBM12A~7, 9, 
10, and 11 is typical of both active field dwarfs and young stars, while 
the much stronger emission from MBM12A~8 is expected for only the latter.
MBM12A~8, 9, and 10 also show IR excess emission in their $J-H$ and $H-K_s$
colors (\S~\ref{sec:disk}). In addition, MBM12A~7 is probably the source
of X-rays detected by H00b at $7\arcsec$ from the optical position of the star. 
Finally, the positions of these five stars 
are closely correlated with those of the original six members, as seen in 
Figure~\ref{fig:map1}. For these reasons, I conclude that MBM12A~7-11
are young stars and are associated with the previously known young 
sources MBM12A~1-6. I used the methods of spectral classification 
for young late-type stars described by Luhman (1999) to derive spectral types
for MBM12A~3, 4, and 6-11, which are listed in Table~\ref{tab:mem}.

\subsection{Evaluation of Remaining Candidate Members}
\label{sec:cand}

Among the stars that were not rejected as field stars in the diagram 
of $R-I$ vs.\ $I$ in Figure~\ref{fig:ri}, there remain 82 sources that 
have not been observed spectroscopically. Are any of these objects promising
candidate members of MBM12A? There are 32 sources at $I<13$ in 
Figure~\ref{fig:ri} that lack spectra. It is unlikely that any of these 
stars are young members because the previous H$\alpha$ and X-ray searches by 
Stephenson (1986) and H00b should be complete at these bright levels. 

The 39 stars at $R-I>2.3$ without IR detections are not easily evaluated. 
From the optical data alone, they could be very cool objects -- foreground 
dwarfs or low-mass members of MBM12A -- or reddened background stars.
Most of these objects are probably an extension of the reddened background star 
population at the left of the boundary in Figure~\ref{fig:ri}. To better 
determine their nature, they must be observed with IR photometry and examined
with the color-color diagrams described in \S~\ref{sec:ident}. If any of them 
are members of MBM12A, they would have masses of $<0.03$~$M_{\odot}$. 

\begin{figure}
\plotone{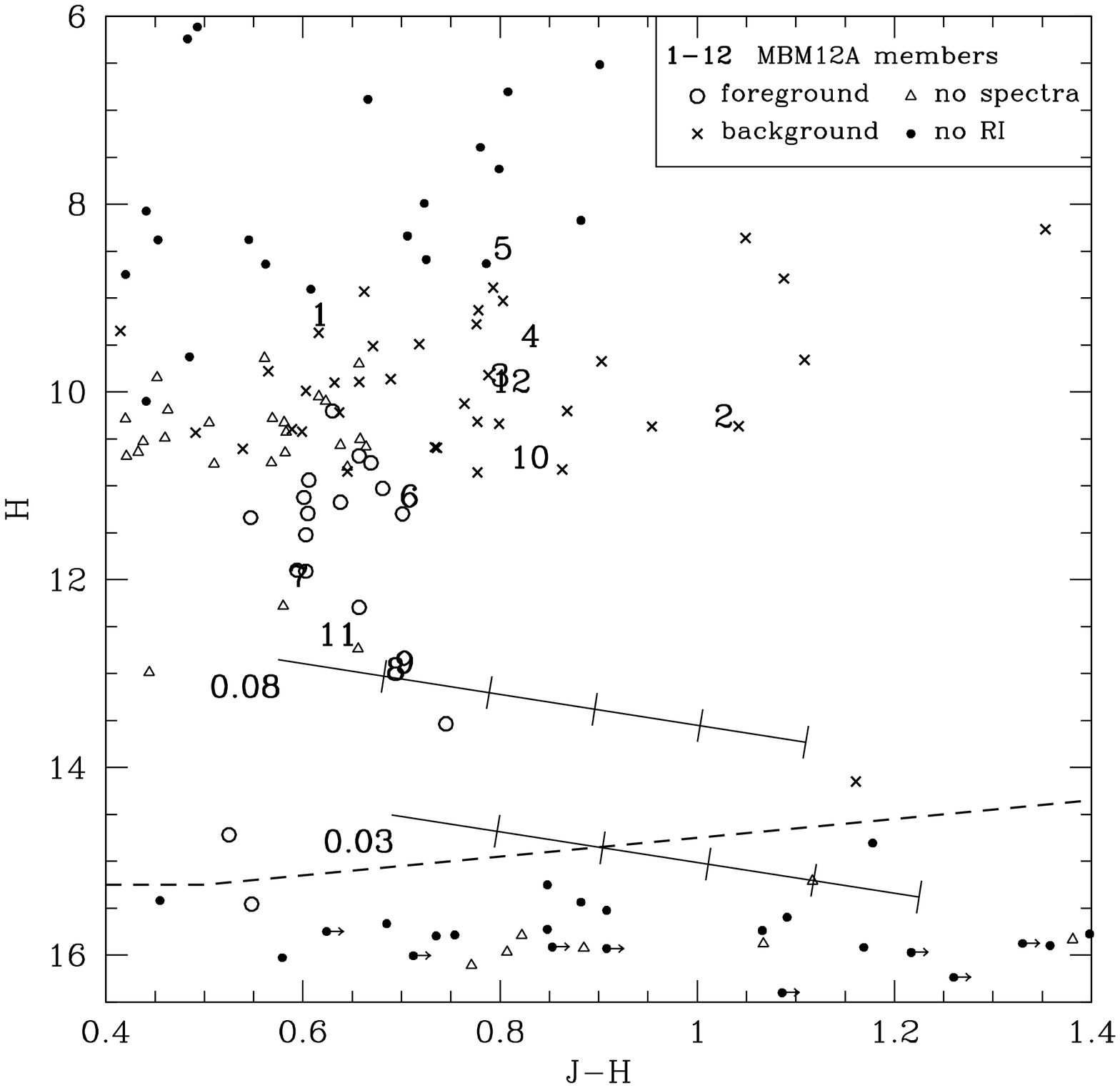}
\caption{
$J-H$ vs.\ $H$ from the 2MASS survey for the MBM12 cloud.
MBM12A~1-6 are previously known young stars.
Spectroscopy of candidate members from $R-I$ vs.\ $I$ in Figure~\ref{fig:ri} 
has been used to identify new members (MBM12A~7-11), foreground dwarfs 
({\it circles}), and background stars ({\it crosses}).
The remaining stars that were not rejected as field stars
in Figure~\ref{fig:ri} and that lack spectra are shown ({\it open triangles}), 
while the probable field stars from that diagram are omitted. 
The young star MBM12A~12 (S18) is located $2.5\arcdeg$ southeast of 
the MBM12 cloud and is a likely member of this association.
The upper and lower lines represent the reddening vectors 
for 0.08 ($\sim$M6.5) and 0.03~$M_{\odot}$ ($\sim$M8) for a distance modulus
of 7.2 (275~pc) and an age of 2~Myr (Baraffe et al.\ 1998).  
The reddening vectors are marked at intervals of $A_V=1$.
The dashed line represents the completeness limits of $J=15.75$ and $H=15.25$.
}
\label{fig:hjh}
\end{figure}

There are 11 objects remaining that lack spectra, all of which have IR 
photometry, $R-I>1.5$, and $I>14$. As described in \S~\ref{sec:ident}, 
at these red $R-I$ colors, young members of MBM12A should be M-type and 
distinguishable from most background stars by diagrams of $J-H$ vs.\ $I-K_s$
and $R-I$ vs.\ $I-K_s$. 
Indeed, the new M5-M6 young stars (MBM12A~7-9, 11) and the three M6-M9 
foreground dwarfs were expected to have M types by their colors in such 
diagrams. Five of the 11 objects that lack spectra do not have the appropriate
colors for late-type stars.  The other six sources are potential members 
of MBM12A, which are listed in Table~\ref{tab:cand} and are discussed
individually.  For 2MASSs~J0252490+195252, 2MASSs~J0258523+195840 and 
2MASSs~J0258344+194650, the colors 
$R-I$, $I-K_s$, $J-H$, and $H-K_s$ are all consistent with spectral types 
of M4, M5, and M5, respectively, with no reddening and little or no IR excess 
(E($J-H$) and E($H-K_s)<0.1$~mag). 
These objects could be either foreground dwarfs or young members of MBM12A. 
The stars 2MASSs~J0255056+194454 and 2MASSs~J0255372+204311 have very similar
photometry, where their optical and IR colors suggest spectral types of M8-M9.
They could be either foreground dwarfs or substellar members of 
MBM12A. If these two stars are members of MBM12A, they should have masses of 
$\sim0.02$~$M_{\odot}$ by their positions in the diagram of 
$J-H$ vs.\ $H$ in Figure~\ref{fig:hjh} ($H\sim16$, $J-H\sim0.8$).
For 2MASSs~J0255452+192526, the various colors are consistent with either a 
highly reddened mid-M dwarf ($A_V=4$-5) or an L-type object with no reddening; 
the observed colors are somewhat better matched by those of a reddened dwarf.
If this source is a member of MBM12A, its position in the diagram of $J-H$ 
vs.\ $H$ in Figure~\ref{fig:hjh} would imply a mass of $\sim0.02$~$M_{\odot}$.
Because 2MASSs~J0255452+192526 is projected against the strongest
peak in the {\it IRAS} 100~$\mu$m image of the MBM12 cloud,
it is probably a highly reddened background dwarf.

\subsection{Completeness}

Because the mass and reddening vectors are roughly perpendicular in a
near-IR color-magnitude diagram and because all of the late-type members 
should have similar intrinsic IR colors, completeness 
in mass and reddening are readily evaluated with near-IR data. 
A diagram of $J-H$ vs.\ $H$ from the 2MASS survey is shown in 
Figure~\ref{fig:hjh} for the field towards the MBM12 cloud that was imaged in 
the $R$ and $I$ bands. The objects that were identified as probable field
stars by the diagram of $R-I$ vs.\ $I$ have been omitted. 
The completeness limits of the 2MASS data are taken to be the magnitudes at 
which the logarithm of the number of sources as a function of magnitude departs 
from a linear slope and begins to turn over ($J\sim15.75$, $H\sim15.25$).
As demonstrated in Figure~\ref{fig:hjh}, the census of stellar members of
MBM12A is complete to the highest line-of-sight reddenings in the cloud 
($A_V=5$-10). All of the known members show reddenings of $A_V<2$ 
(\S~\ref{sec:ext}). If the range of reddenings is the same for stellar
and substellar members, then the 2MASS completeness limit indicates that 
this search for new members of MBM12A is complete to 0.03~$M_{\odot}$, 
with the exception of the three candidates at $H=12$-13 from 
Table~\ref{tab:cand}.
The only other IR sources for which membership status is unknown
are the bright sources at $H<11$ -- which are unlikely to be members 
(\S~\ref{sec:cand}) -- and the stars near the 2MASS detection limit -- which 
should have masses of $\sim0.02$~$M_{\odot}$ if they are members.

\subsection{RXJ0306.5+1921 and MBM12A~12}
\label{sec:1921}

In a search for new members of MBM12A, H00b used the presence
of X-ray emission to identify candidates and obtained followup spectroscopy
to check for signatures of youth such as Li absorption and H$\alpha$ emission.
They found that the star RXJ0306.5+1921, which is $\sim2\arcdeg$ east of the
cloud, exhibited Li absorption (0.35~\AA) and a spectral type of K1 in a
low-resolution spectrum (6.4~\AA). 
Because of the X-ray emission and moderately strong Li absorption, H00b 
identified RXJ0306.5+1921 as a weak-line T Tauri star that was likely a member
of MBM12A. However, as discussed in \S~\ref{sec:sptypes}, moderately high 
spectral resolution
($<2$~\AA) is necessary for accurately measuring the strength of Li absorption 
in early K stars. Insufficient resolution can produce overestimates of the 
line strength because of line blending. Indeed, Jayawardhana et al.\ (2001) 
recently measured a Li equivalent width of only 0.22~\AA\ in higher 
resolution spectra (1.5~\AA) of RXJ0306.5+1921. This Li strength is
more indicative of a star at an age of 100~Myr than a T~Tauri 
star (\S~\ref{sec:li}). Furthermore, RXJ0306.5+1921 lacks any other 
distinguishing characteristics of young stars. The star shows no excess 
emission at near- to mid-IR wavelengths (Jayawardhana et al.\ 2001), no
reddening, and little or no H$\alpha$ emission. 
Finally, when placed on the H-R diagram at the 
distance to MBM12A, RXJ0306.5+1921 does not have the same age as the members 
of MBM12A (\S~\ref{sec:hr}). There is no evidence that RXJ0306.5+1921 is a 
young star or a member of MBM12A.

As with RXJ0306.5+1921, the star MBM12A~12, also known as S18, has been taken 
to be a member of MBM12A although it is offset from the cloud by a few degrees. 
The IR excess emission, Li absorption, and H$\alpha$ emission for MBM12A~12 
are clear evidence that it is a bonafide T~Tauri star (H00b; Jayawardhana et 
al.\ 2001).  In addition, the age of MBM12A~12 on the H-R diagram is 
consistent with that of the association when all stars are placed at the same 
distance (\S~\ref{sec:hr}). It is likely that MBM12A~12 was produced 
in the same episode of star formation as the young stars projected against
the MBM12 cloud, but proper motion or radial velocity data are needed to 
determine its membership in the association with better certainty. 
For the purposes of this study, MBM12A~12 is taken to be a member of MBM12A.
 
\section{The MBM12A Stellar Population}

\subsection{Distance}
\label{sec:distance}

\subsubsection{Using the Cloud to Derive the Distance of the Association}

To estimate the distance to MBM12A,
I first consider whether the association and the cloud are at the same distance.
As shown in Figure~\ref{fig:map1}, with the exception of MBM12A~12, the 
positions of the young stars are closely correlated with the 
{\it IRAS}~100~$\mu$m emission from the cloud.
In addition, radial velocities have been measured for two members of the
association, MBM12A~1 and 4, and are consistent with the
distribution of velocities of the cloud (H00b). 
The reddenings of the stars ($A_V=0$-2) are less than those observed for
background stars at comparable sight lines ($A_V=3$-8), which clearly
indicates that the stars are not behind the cloud. It is likely that the young 
stars projected against the MBM12 cloud are at the same distance as the cloud. 

\subsubsection{Cloud Distance from Interstellar Na~I Absorption}

The commonly referenced distance to the MBM12 cloud is 65~pc as reported
by Hobbs et al.\ (1986). They obtained high-resolution spectroscopy 
of the Na~I D lines in A and F stars along the line of sight to the cloud
and identified probable foreground and background stars by the absence 
and presence of interstellar absorption in these transitions. This
information was then combined with spectroscopic distances to constrain 
the distance to the cloud. From the apparent foreground and background 
positions of HIP13631 and $\epsilon$~Ari~AB (HIP13914),
the cloud distance was estimated 
to be 60-70~pc, which would make it one of the nearest molecular clouds.
Hearty et al.\ (2000a) updated this analysis by including Na spectroscopy of
additional stars near MBM12 and by replacing the spectroscopic distances with 
{\it Hipparcos} trigonometric measurements. In the end, the distance estimate
of Hearty et al.\ (2000a) was based on the same pair of stars as in the study
by Hobbs et al.\ (1986), HIP13631 ($58\pm5$~pc) and $\epsilon$~Ari~AB 
($90\pm12$~pc). 
However, there are two additional stars in the combined samples of
Hobbs et al.\ (1986) and Hearty et al.\ (2000a) that provide 
useful limits on the distance to the cloud. As shown in Figure~\ref{fig:map2}, 
the stars HIP13913 ($91\pm8$~pc) and HIP13855 ($74\pm8$~pc)
are projected against levels of {\it IRAS}~100~$\mu$m emission that 
are comparable to those at the positions of HIP13631 and $\epsilon$~Ari~AB,
making them equally valid stars for constraining the cloud distance. 
By their proximity to the cloud on the sky and their lack of Na absorption, the 
stars HIP13834 ($32\pm1$~pc), HIP13631 ($58\pm5$~pc), HIP13855 ($74\pm8$~pc),
and HIP13913 ($91\pm8$~pc) are probably in the foreground of the cloud and 
therefore imply a distance of $\gtrsim90$~pc to MBM12.

I now discuss the star $\epsilon$~Ari~AB, which has previously provided the 
upper limit of $\sim90$~pc for the distance to MBM12.
If this star is behind the MBM12 cloud, then it should exhibit 
reddening in its colors, but none is apparent in $B-V$ and $V-I$.
This discrepancy was noted by Hobbs et al.\ (1988), who suggested 
that the intrinsic colors used in computing the reddening may be uncertain 
due to the multiplicity of the system. In other words, is it possible
that the system does in fact suffer from reddening, but that its intrinsic 
colors are bluer than the assumed values? Spectral types of A2IV and A3IVs
have been measured for the individual components of $\epsilon$~Ari~AB by
Abt \& Morrell (1995) and colors of $B-V=0.048$ and $V-I=0.05$
are listed in the {\it Hipparcos} database for the system. 
In the following discussion, the A2IV primary is assumed to dominate the 
photometry. If the secondary did contribute significantly to the colors, 
the intrinsic colors of the system would be redder and make it even less
plausible that the system is reddened.
The observed colors of the system agree perfectly with 
those expected for a spectral type of A2V. 
As an additional test, I consider the 49 other stars classified as A2IV in 
the study of Abt \& Morrell (1995), which should have colors that are very 
similar to those of the primary component of $\epsilon$~Ari~AB. For those
49 stars, the average color is $B-V=0.064$ and the bluest value is $B-V=0.01$. 
Given this typical color, $\epsilon$~Ari~AB again exhibits no reddening. 
Even if it has colors that are the same as the bluest stars in that sample, 
the reddening would be only $A_V=0.1$.
In conclusion, there is no evidence of the reddening towards $\epsilon$~Ari~AB 
that is expected if the interstellar Na absorption in its spectrum arises 
from a dark cloud in the foreground of the star.

Hobbs et al.\ (1986) did note the possibility that an unusually 
high concentration of foreground neutral gas could be responsible for the
Na absorption towards $\epsilon$~Ari~AB and that the MBM12 cloud could fall at
some distance behind the star, which would explain the lack of 
reddening towards the star. They considered this an unlikely possibility for
the following reasons.
The Sun is surrounded by a cavity containing hot ionized gas that extends out
to $\sim100$~pc, which is known as the local bubble (Cox \& Reynolds 1987).  
Because the spectroscopic 
distance of 70~pc used by Hobbs et al.\ (1986) for $\epsilon$~Ari~AB placed 
the star within this cavity, it appeared unlikely that a neutral gas cloud 
of high optical depth would fall in the foreground of the star. 
However, the {\it Hipparcos} distance to $\epsilon$~Ari~AB is $90\pm12$~pc,
which places it near the wall of the local bubble. In fact, in a study 
where {\it Hipparcos} distances were combined with Na measurements to map
the structure of the local bubble, Sfeir et al.\ (1999)
attributed the Na absorption for $\epsilon$~Ari~AB to a clump of cold neutral 
interstellar gas that is possibly associated with the local bubble wall.
A similar scenario was suggested for HD14613 and HD14670, which also have
strong Na absorption and are near a clumpy portion of the wall of the local
bubble. As with $\epsilon$~Ari~AB, these stars exhibit no reddening in their 
$B-V$ colors. Given this explanation for the Na absorption and the lack of 
reddening, it is likely that $\epsilon$~Ari~AB and the wall of the local 
bubble are in the foreground of the MBM12 cloud. As a result,
this star cannot be used to provide the upper limit to the distance to
the cloud. In fact, because the boundary of the local bubble is mapped by
searching for the Na absorption that is expected from the interstellar medium
beyond the bubble, the presence of Na absorption cannot be used as evidence
that a star is behind a dark cloud if that star is beyond the local bubble,
or at distances greater than 90~pc for the direction of MBM12. In other words,
the Na absorption detected for the stars beyond 90~pc in
Figure~\ref{fig:hip} could be attributed to the interstellar medium rather
than the MBM12 cloud.
Therefore, the Na absorption measurements provide only a lower limit of 
$\sim90$~pc for the distance to MBM12.

\subsubsection{Cloud Distance from Reddening}

While Na absorption can be produced by both molecular clouds and the neutral 
gas in the interstellar medium, reddening is by definition a direct tracer of 
a dark cloud. I explore two methods that incorporate reddening in measuring
the distance to MBM12.

\subsubsubsection{Stars with Trigonometric Parallaxes}

\begin{figure}
\plotone{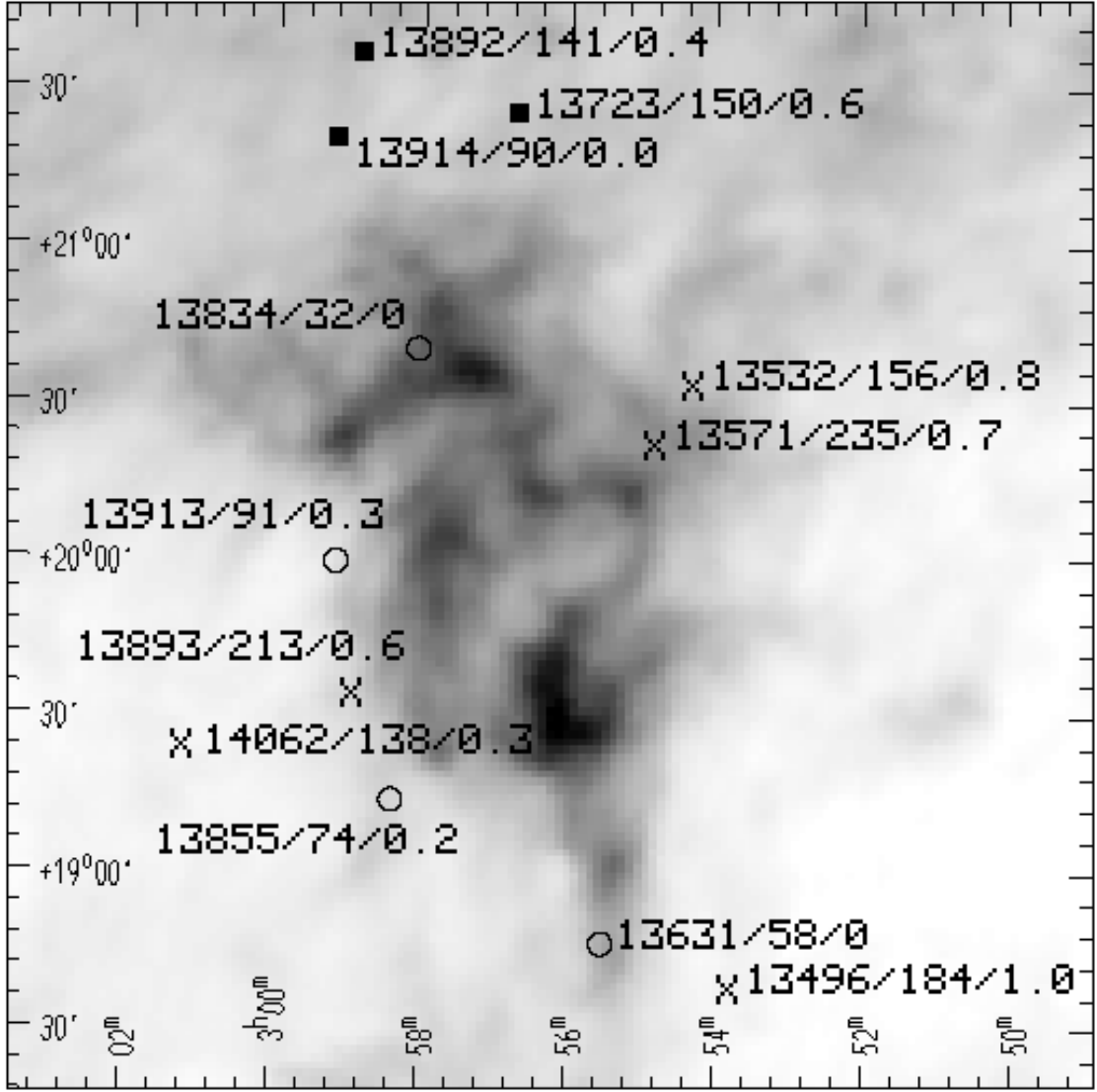}
\caption{
The {\it IRAS} 100~$\mu$m image of the MBM12 cloud is shown with the 
positions of all stars that are projected near the cloud and that have 
{\it Hipparcos} distances, where the symbols indicate 
the presence or absence of interstellar Na~I absorption ({\it solid squares
and open circles}) or the lack of a measurement ({\it crosses}).
Each star is labeled with its HIP designation, distance (pc), and $A_V$.
}
\label{fig:map2}
\end{figure}

\begin{figure}
\plotone{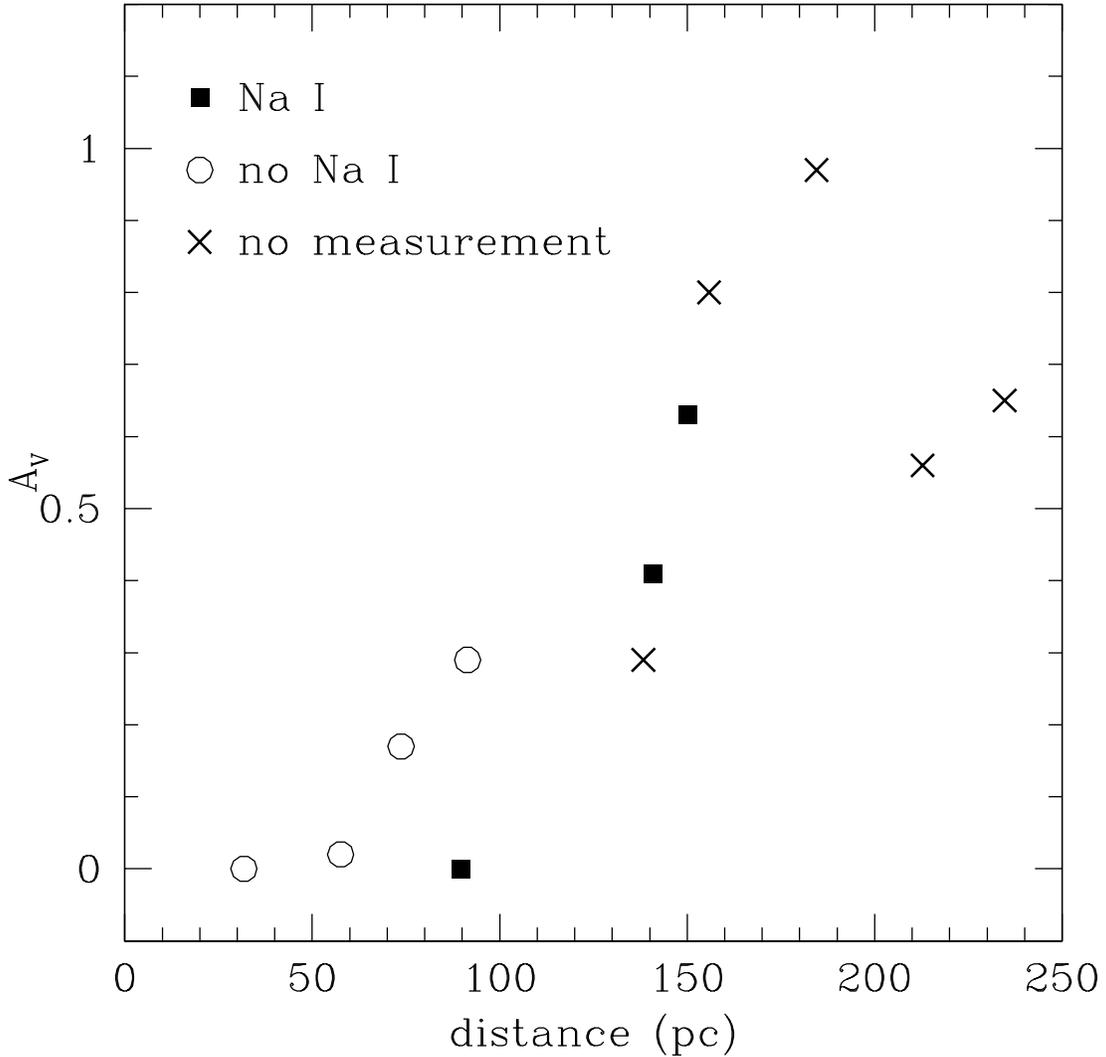}
\caption{
Distance vs.\ $A_V$ for stars that are projected near the MBM12 cloud and that
have {\it Hipparcos} distances, where the symbols indicate the presence or 
absence of interstellar Na~I absorption ({\it solid squares and open circles}) 
or the lack of a measurement ({\it crosses}). The stars showing no Na absorption
are probably in front of the cloud. It is unclear whether the stars
with Na absorption are in the foreground or background of MBM12 because
the absorption could arise from either the cloud or the interstellar medium
beyond the local bubble. Meanwhile, none of these stars show the elevated
reddenings ($A_V>1$) expected if they were behind the cloud; however, because
the more distant stars are on the outskirts of the cloud, 
they could be behind the cloud and yet have fairly low reddenings. 
Thus, for the distance to the MBM12 cloud, the Na data indicate a lower limit 
of 90~pc and the reddenings are consistent with a lower limit of 200-250~pc. 
}
\label{fig:hip}
\end{figure}

For the stars near the cloud that have {\it Hipparcos} distances,
their foreground or background nature can be examined through their reddenings.
The positions of the 12 stars in the immediate vicinity of the MBM12 cloud on 
the sky that have {\it Hipparcos} distances are plotted in 
Figure~\ref{fig:map2}. Colors in $B-V$ are taken from the {\it Hipparcos} 
database. Data at $K_s$ are available from the 2MASS survey for 7 of these 
stars, which are combined with the $V$ measurements compiled by {\it Hipparcos} 
to arrive at $V-K_s$ colors. 
An approximate luminosity class (dwarf or giant) of 
each star was inferred from its absolute magnitude. 
Extinctions are calculated from the $B-V$ and $V-K_s$ colors by assuming the
appropriate intrinsic values expected for the spectral type and luminosity 
class in question (see \S~\ref{sec:ext}). 
The reddenings and distances for the 12 stars are shown on the map in
Figure~\ref{fig:map2} and are used to construct the diagram of distance
vs.\ reddening in Figure~\ref{fig:hip}. 
The reddenings among these stars increase with distance in a manner that is
consistent with the reddening arising from the interstellar medium; there is
no abrupt increase in reddening that is expected if any of these stars 
are behind the MBM12 cloud. The lack of significant reddening
towards these stars implies
that the cloud has a distance of more than 200-250~pc. However, because 
the more distant stars in this sample are on the outskirts of the cloud, 
they could be behind the cloud and yet have fairly low reddenings. 

\subsubsubsection{Foreground and Background M dwarfs}
\label{sec:dwarfs}

I now estimate the distance to the MBM12 cloud by using IR and optical 
photometry to identify background and foreground dwarf stars and then comparing 
their apparent magnitudes to the absolute magnitudes of nearby dwarfs at
comparable spectral types.

\begin{figure}
\epsscale{0.75}
\plotone{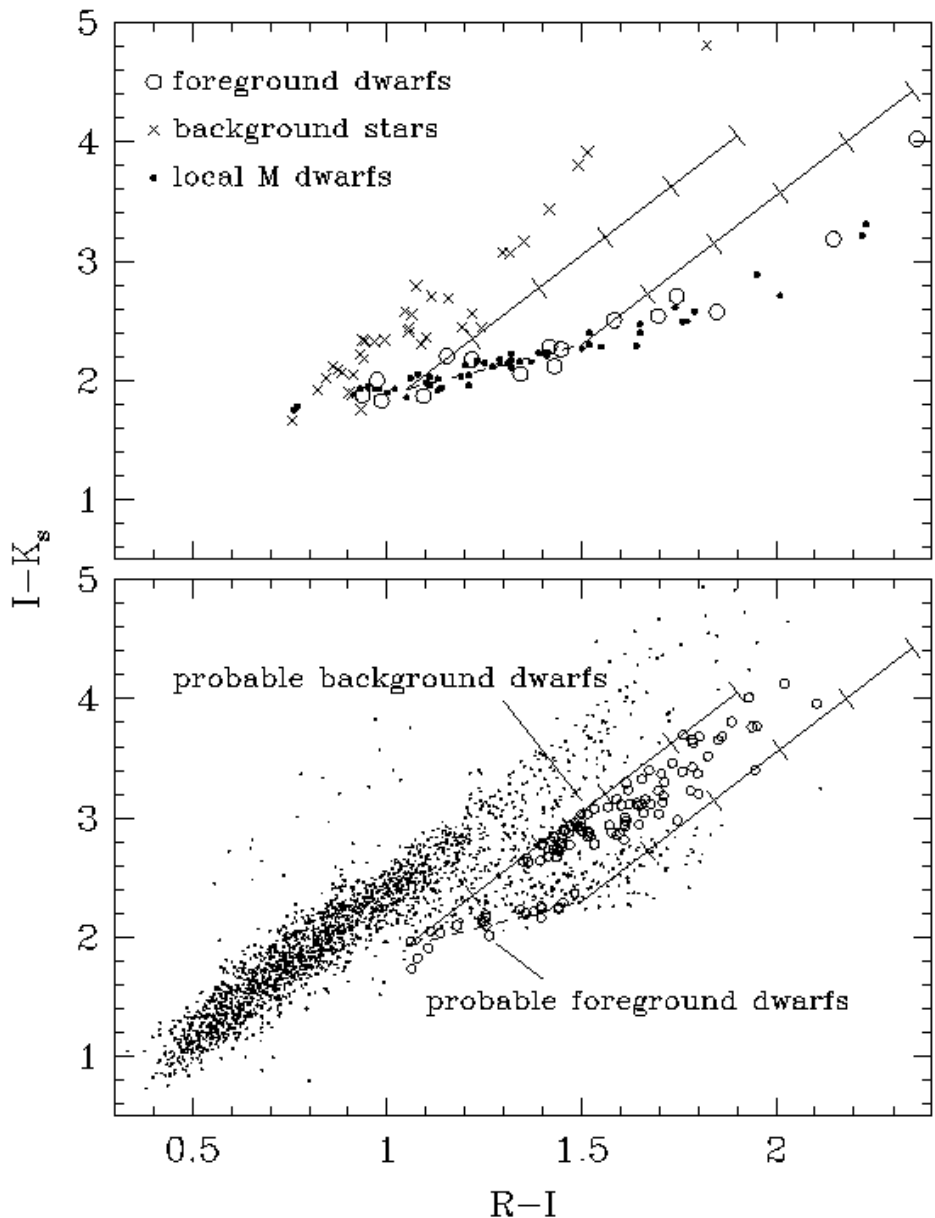}
\caption{
In the upper panel, $R-I$ vs.\ $I-K_s$ is shown for the spectroscopically 
identified foreground dwarfs ({\it open circles}) and background giants and 
early-type stars ({\it crosses}) towards MBM12 and for local M dwarfs 
({\it solid points}). A fit to the colors of M dwarfs between $R-I=1.05$ and 
1.5 is plotted ({\it dashed line}). Reddening vectors originate from 
each end of the fit and are marked at intervals of $A_V=1$ ({\it solid lines}).
In the lower panel, probable foreground and background stars 
({\it open circles}) are identified among stars towards MBM12 that lack spectra.
Stars that are near the M dwarf fit and are projected against the higher 
levels of {\it IRAS} 100~$\mu$m emission of the cloud are labeled as likely 
foreground M dwarfs. 
Stars in the reddening band for the M dwarf fit with colors indicative of
$A_V>1.5$ are likely to be M dwarfs behind the MBM12 cloud.
}
\label{fig:riik}
\end{figure}

For this method to be successful, there must be a range of dwarf spectral types
that is distinguishable from all other types through photometry alone, even
in the presence of variable reddening. 
Because of reddening, stars of different spectral types can inhabit the 
same regions of most color-color and color-magnitude diagrams. 
However, in a diagram of $R-I$ vs.\ $I-K_s$, M dwarfs have positions that are 
unique from those of other stars, as demonstrated in the top panel of 
Figure~\ref{fig:riik} where the colors of the foreground dwarfs and the 
background giants and early-type stars from the spectroscopic sample are 
compared.
(This diagram could have been used to better filter the candidate members that 
were observed with spectroscopy, as noted in \S~\ref{sec:ident}).
At intrinsic colors of $R-I\gtrsim1.05$ ($\gtrsim$M1), M dwarfs become 
well-separated from other stars. Because the path of the latest M dwarfs 
becomes nearly parallel to the reddening vector in $R-I$ vs.\ $I-K_s$, their
reddenings cannot be accurately estimated from this combination of colors. 
On the other hand, the extinctions for dwarfs at $R-I\lesssim1.5$ can be readily
determined by dereddening the $R-I$ and $I-K_s$ colors to the sequence of
intrinsic colors represented by the dashed line in Figure~\ref{fig:riik}.
Therefore, through $R-I$ vs.\ $I-K_s$, dwarfs from $R-I=1.05$ to 1.5 (M1-M4)
can be separated from stars of other spectral types and have their reddenings 
estimated.
The lower panel of Figure~\ref{fig:riik} shows all stars in the images of
MBM12 that lack spectra and have $H<15$ and $I<19$. 
Stars that are near the M dwarf fit at $R-I=1.05$-1.5 and that are projected 
against the higher levels of {\it IRAS} 100~$\mu$m emission are likely to be
M dwarfs in the foreground of the cloud.
The remaining stars within the reddening 
band for the M dwarf fit are probably reddened M dwarfs, where those with 
colors indicative of $A_V>1.5$ are very likely to be behind the MBM12 cloud. 

\begin{figure}
\plotone{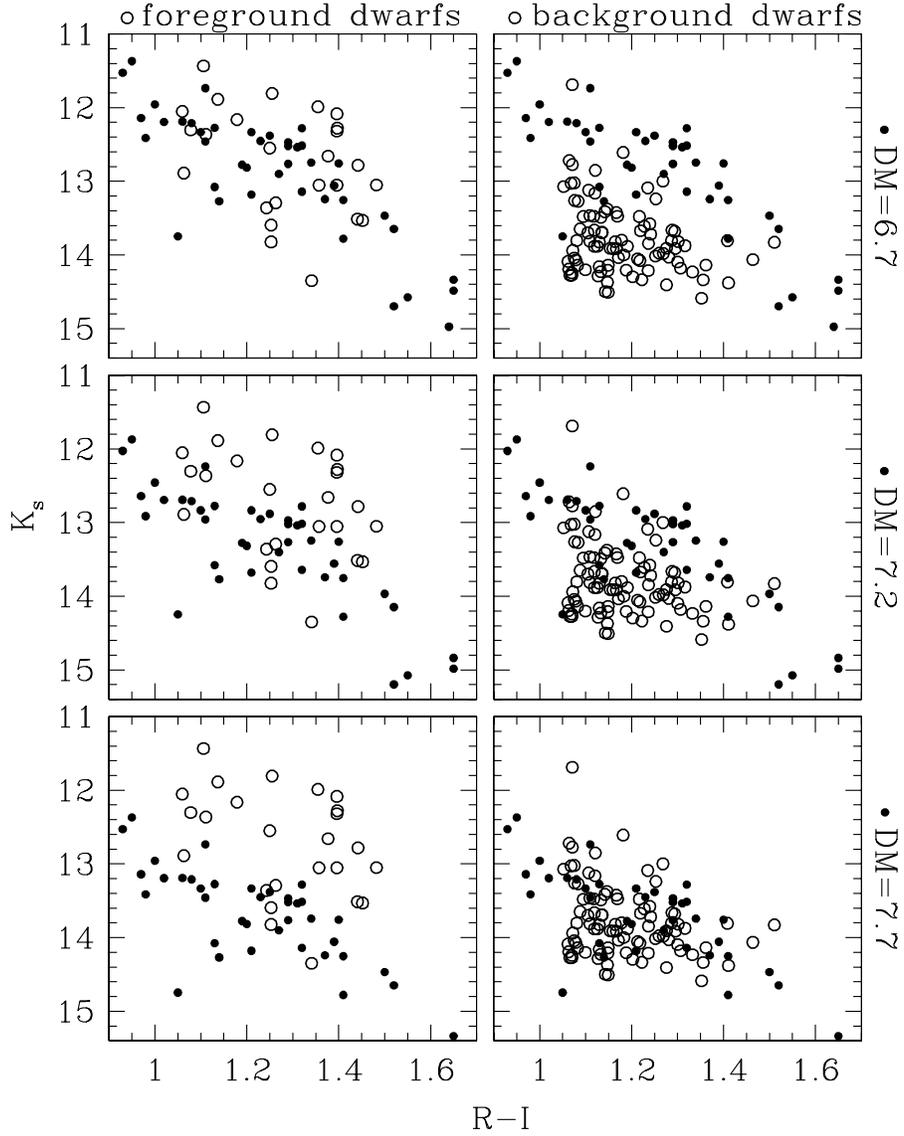}
\caption{
The distance to the MBM12 cloud is inferred from these diagrams of
$R-I$ vs.\ $I$. 
Stars identified as likely foreground and background M dwarfs 
by Figure~\ref{fig:riik} are shown on the left and right ({\it open circles}).
The data for the background dwarfs have been corrected for reddening. 
For comparison, local M dwarfs are placed at three distances
from top to bottom ({\it solid points}). 
A comparison of the lower and upper envelopes of the M dwarf sequence to 
those of the foreground and background dwarf distributions, respectively, 
indicates a distance modulus of $7.2\pm0.5$ (275~pc) to the MBM12 cloud.
}
\label{fig:rik}
\end{figure}

The two samples of probable foreground and background M dwarfs can be
compared to local M dwarfs in a color-magnitude diagram to constrain the 
distance to the MBM12 cloud. For this diagram, $R-I$ is selected as the color
because it changes significantly from types of M1 to M4. The 2MASS measurements
at $K_s$ are used for the magnitude; the correction for reddening is small
in this band and the availability of 2MASS data for both the MBM12 stars and 
a large number of local M dwarfs enables a reliable comparison among all stars.
The foreground and background dwarfs towards MBM12 are shown in the 
diagrams of $R-I$ vs.\ $K_s$ in Figure~\ref{fig:rik}. The data for the
background dwarfs have been corrected for the reddenings implied by 
Figure~\ref{fig:riik}. The local field dwarf sample consists of all 
single M stars from the NASA NStars database ($<25$~pc) for which $R-I$ and
2MASS $K_s$ data are available. The individual distances and $R-I$ colors
compiled by NStars and the $K_s$ measurements from 2MASS are used to place 
the stars in this local dwarf sample at distance moduli of 6.7, 7.2, and 7.7 
in Figure~\ref{fig:rik}. If the field dwarfs in the vicinity of MBM12
and those in the solar neighborhood have similar photometric properties, 
and if the latter are placed at the distance to MBM12, 
than the lower envelope of the distribution of dwarfs in the foreground of 
MBM12 should match the lower envelope of the local M dwarf sequence.
At the same time, the upper envelopes of the background dwarfs and the
local M dwarfs should agree as well. As illustrated in Figure~\ref{fig:rik},
this method indicates a distance modulus of $7.2\pm0.5$ ($\sim275$~pc) to the 
MBM12 cloud.

\subsubsection{Implications of New Distance}

The revision in the distance to MBM12 from 65 to 275~pc has implications for 
the previously derived properties of both the cloud and the local bubble.
With a distance of 65~pc, Zimmerman \& Ungerechts (1990)
found that the cloud is not gravitationally bound, except perhaps for
a few of the larger clumps. Although the larger revised distance suggests
the cloud is more easily bound, these general conclusions may remain unchanged
with the larger distance (Zimmerman \& Ungerechts 1990).
Because of its shadow against the diffuse soft X-ray background and its
suggested distance of 65~pc, the MBM12 cloud was believed to fall within 
the local bubble (Snowden, McCammon, \& Verter 1993). 
In estimating the size of the local bubble from X-ray data, 
Snowden et al.\ (1998) assumed a distance intermediate between 32 to 90~pc 
to the MBM12 cloud. More recently, 
Sfeir et al.\ (1999) found that the boundaries of the local bubble 
as inferred from those X-ray measurements and from Na data agree better
if MBM12 is placed at or beyond the edge of the local
bubble ($\gtrsim90$~pc) in the X-ray analysis, which is consistent with
the revised distance of 275~pc derived in this work.
As a result, the X-ray shadow towards the MBM12 cloud is probably
caused by diffuse emission from a region outside of the local bubble, such as 
the Galactic Halo, as observed for other distant clouds (Snowden et al.\ 2000).

\subsection{Age}
\label{sec:age}

The age of MBM12A is constrained through two independent methods in the
following discussion. First, its age relative to other young associations 
is examined through a comparison of Li strengths between the populations. 
Relative and absolute ages are then inferred from the positions of these
associations on the H-R diagram. For the comparison to MBM12A, I have
selected four of the nearest associations and clusters that have ages of 
10-30~Myr: $\eta$~Cha, TWA, HorA, and IC~2602. 
The sources of the data for these young
associations are described in detail in the appendix.

\subsubsection{Lithium}
\label{sec:li}

As young low-mass stars evolve from the birthline ($\sim1$~Myr) to the main 
sequence ($\sim100$~Myr), the processes of convection and nuclear burning 
deplete Li at their stellar surfaces. Consequently, the age of a 
coeval population of stars is reflected in the typical strengths of 
photospheric Li absorption in those stars.

The behavior of Li absorption is dependent on both age and effective 
temperature.  At M spectral types, the Li equivalent widths vary from 
0.45-0.8~\AA\ at $\sim1$~Myr (e.g., Taurus; Basri et al.\ 1991; 
Mart{\'\i}n et al.\ 1994) to nearly zero at
$\gtrsim30$~Myr (e.g., IC~2602; Randich et al.\ 1997). For G and K stars, 
the initial Li absorption is weaker and the species is not fully depleted even 
at 100~Myr, resulting in a smaller change in observed Li strengths with time.
Consequently, the diagram of $T_{\rm eff}$ vs.\ $W_{\lambda}$(Li) in
Figure~\ref{fig:li} is useful for comparing the Li strengths among MBM12A 
and the other young populations. The distributions of Li strengths for
MBM12A, $\eta$~Cha, and TWA are indistinguishable from each other.
The range of equivalent widths (0.4-0.65~\AA) for the M stars in these 
associations is similar to or slightly less than that observed for Taurus.
Meanwhile, the Li data clearly indicate that
MBM12A, $\eta$~Cha, and TWA are younger than IC~2602 and HorA. 
The Li measurements are consistent with the same age for the latter two 
populations. However, because the change in Li strengths is small from
IC~2602 (30~Myr) to the Pleiades (125~Myr) and the number statistics are poor
for HorA, these Li data also allow for an age difference of 
few tens of millions of years between IC~2602 and HorA. 

Because H00b measured strong Li absorption for RXJ0306.5+1921,
they identified it as a young star and thus a possible 
member of MBM12A. However, the weaker line strength found at 
higher spectral resolution (Jayawardhana et al.\ 2001) implies
an age of $\gtrsim100$~Myr, supporting the assertion that this star
is not a young member of MBM12A (\S~\ref{sec:1921}).

\subsubsection{H-R Diagram}
\label{sec:hr}

In constructing the H-R diagrams for the young associations in 
Figure~\ref{fig:hr}, the luminosities and temperatures have been 
derived in the most uniform fashion that is possible given available data.
By then placing these associations on one set of theoretical evolutionary 
models (Baraffe et al.\ 1998), their relative ages can be determined from a
consistent and meaningful comparison among the populations. 

On an H-R diagram, a very young stellar population in a star forming region 
($\lesssim1$~Myr) will form a locus that has a large range
in luminosities at a given temperature. The dominant sources of this width
in the locus are uncertainties in luminosity estimates, variability,
and a spread in ages among the cluster members. As a result, when using
evolutionary models to infer an age for the population, it is the median
or mean age that is generally measured and reported. For a stellar cluster
that is older than 5-10~Myr, the lack of reddening and circumstellar excess 
emission allows more accurate luminosity estimates, accretion-related
variability is not present, and typical age spreads of a few million years
correspond to negligible ranges of luminosities. Instead, the largest 
contributor to the width of the locus for an older cluster is expected to
be unresolved binaries that appear as overly luminous single stars. Therefore,
for older populations the age is measured from the sequence of single stars 
that forms the lower envelope of the locus rather than from the median of the 
locus. This approach is clearly applicable to IC~2602. 
Because of the lack of excess emission near the bands from which the 
luminosities are computed and the low level of reddening for stars in 
$\eta$~Cha and MBM12A, unresolved binarity could be the source of the 
spread in luminosities of their sequences as well. 
In fact, the presence of double H$\alpha$ emission profiles for some of the 
brighter members of MBM12A and $\eta$~Cha suggests that they may be binaries
(Jayawardhana et al.\ 2001; Mamajek et al.\ 1999). Therefore, the ages
for these associations will be inferred from the lower envelopes of their
sequences. For TWA and HorA, the large uncertainties in the relative distances 
among their members may dominate the scatter in the sequences. 
Therefore, the age of TWA will be measured from the position of TW~Hya alone
and the age of HorA will be estimated by considering both the median and the 
lower envelope of its sequence.

Previous studies have estimated distances of 50-100~pc to the MBM12 cloud.
As shown in Figure~\ref{fig:hr}, this range of distances implies that the
age of MBM12A is equal to or greater than that of IC~2602 and HorA. 
If so, the Li strengths of MBM12A should be similar to 
those of IC~2602 and HorA, which is clearly not the case
as seen in the previous section. If the commonly accepted distance of 
65~pc were adopted for MBM12A (Hobbs et al.\ 1986), these stars would fall 
at or below the main sequence ($\gtrsim100$~Myr).

For comparison to the other associations, MBM12A is plotted for the distance 
modulus of 7.2 that was derived in \S~\ref{sec:distance}.
A well-defined locus of stars is delineated for both MBM12A and $\eta$~Cha.
As described above, unresolved binaries are the likely source of the 
spread in luminosities for each locus. 
With the binarity effects taken into account, the sequences for MBM12A
and $\eta$~Cha imply ages of 2~Myr and 10~Myr. The uncertainty of $\pm0.5$
in the distance modulus to MBM12A corresponds to possible ages of 1-5~Myr.
A comparison of the disk properties of these three populations suggests 
that IC~348 and MBM12A have similar ages and are slightly older than Taurus,
which supports an age of 2~Myr for MBM12A.
The position of the star TW~Hya on the H-R diagram indicates an age of 10~Myr 
for TWA.

The models of Baraffe et al.\ (1998) suggest an age of $30\pm5$~Myr 
for IC~2602, which agrees with the value reported by Stauffer et al.\ (1997) 
using the calculations of D'Antona \& Mazzitelli (1994). 
For HorA, the members have distances ranging from 40 to 100~pc, 
which is equivalent to variations of $\pm0.4$ in log~$L_{\rm bol}$. 
Because only 4 members of the association have {\it Hipparcos} distances, 
large uncertainties are possible in the distances for most of the members,
which could result in errors in the luminosity estimates that are 
greater than the effect of unresolved binaries. Consequently, the age
of the association may be reflected in either the median or the lower envelope
the sequence, which imply ages of $\sim30$ and 50~Myr, respectively.
An age of $30^{+20}_{-5}$~Myr is adopted for HorA. 
These relative and absolute ages for IC~2602 and HorA are consistent 
with the age constraints provided by the Li data.

The position of RXJ0306.5+1921 is plotted with the members of MBM12A in 
Figure~\ref{fig:hr}. As found with the Li data, this star appears significantly 
older than the stars in MBM12A when the same distance is adopted for all stars.

\subsection{Disk Properties}
\label{sec:disk}

The properties of inner disks as reflected by near-IR excess emission 
are compared among MBM12A, Taurus, and the cluster IC~348. 

In Figs.~\ref{fig:jhhk} and \ref{fig:jhkl}, the diagrams of $H-K_s$ 
vs.\ $J-H$ and $K_s-L$ vs.\ $J-H$ are shown for the 12 members of MBM12A and 
for the seven original members that were observed at $L$ by Jayawardhana et 
al.\ (2001), respectively. 
To estimate the intrinsic IR colors of the star-disk systems, these
colors have been corrected for the reddenings in Table~\ref{tab:mem}, most of 
which were derived from $R-I$. 
No optical colors were available for MBM12A~1, 5, and 12, which were 
dereddened by assuming values for the intrinsic IR colors (\S~\ref{sec:ext}).
The locus of intrinsic colors for classical T~Tauri stars (CTTS) in Taurus is 
also shown (Meyer, Calvet, \& Hillenbrand 1997). The origin of the Taurus
locus is near the intrinsic colors of a M0 dwarf, which is the typical 
spectral class of the Taurus young stars from which the CTTS locus was measured.
For other spectral types, the CTTS locus is expected to maintain the same
slope but with the origin shifted to the dwarf colors of the spectral type
in question (Meyer et al.\ 1997; Luhman 1999).

The frequencies and sizes of excess emission in both $JHK_s$ and $JHK_sL$ in
MBM12A are very similar to those in IC~348 (see Figure~1 of Haisch et 
al.\ 2001). Stars in Taurus exhibit greater $K$-band excess emission than
these two regions; the $K$-band excess frequency is $49\pm6\%$ in Taurus as
compared to $18\pm4\%$ in IC~348 (Haisch et al.\ 2000, 2001) and 
the $K$-band excesses of typical CTTS in Taurus are higher than those 
in MBM12A (Figure~\ref{fig:jhhk}).
Meanwhile, as shown in Figure~\ref{fig:jhkl}, the sizes of the $L$-band 
excesses in the MBM12A stars are comparable to those in Taurus. 
In fact, the frequencies of $L$-band excess sources are similar among 
MBM12A (5/7, $71\%\pm32$), Taurus ($69\pm7\%$), and IC~348 ($65\pm8\%$) 
(Haisch et al.\ 2000, 2001; Kenyon \& Hartmann 1995).
This reduction in short-wavelength excess emission with age from Taurus to 
IC~348 and MBM12A (and TW~Hya; Jayawardhana et al.\ 1999) may be a result of 
inside-out clearing or settling of the disk. 

\section{Conclusion}

I have conducted a comprehensive study of the group of young
stars that is projected against the MBM12 dark cloud. 
For this association, I have searched for new members down to substellar
masses, revised its distance, and examined its age and disk properties
in the context of other young populations. 
The conclusions are as follows:

\begin{enumerate}

\item
With IR photometry from the 2MASS survey and new optical imaging and 
spectroscopy, I have performed a census of the MBM12A membership that is 
complete to 0.03~$M_{\odot}$ for a $1.75\arcdeg\times1.4\arcdeg$ field 
towards the MBM12 cloud, from which I find five new members with spectral 
types of M3-M6 (0.1-0.4~$M_{\odot}$). If the IMF in MBM12A were the same as 
that in the Trapezium Cluster (Luhman et al.\ 2000), 2-3 brown dwarfs 
($>0.03$~$M_{\odot}$) are expected in MBM12A where none were found. 
Although this difference between MBM12A and the Trapezium is not significant 
by itself, it is consistent with the deficit of brown dwarfs observed
for Taurus (Luhman 2000), which has similar star-forming conditions as MBM12.
Meanwhile, the previously suggested member RXJ0306.5+1921 shows no evidence 
for youth or membership in the association.
There are a total of 11 young sources that are resolved in normal ground-based
imaging ($\gtrsim1\arcsec$) towards the MBM12 cloud and an additional star
$2\arcdeg$ away towards MBM13 that is probably associated with MBM12A.
I identify a few additional candidate members of MBM12A that 
have not been observed spectroscopically.

\item
I have used optical and IR photometry to identify stars that are likely to
be M dwarfs in the foreground and background of the MBM12 cloud.
By comparing the magnitudes of these stars to those of local field dwarfs, 
the distance modulus to the cloud is constrained to be $7.2\pm0.5$ (275~pc).
MBM12 is not the nearest molecular cloud and is not inside the local bubble
of hot ionized gas as had been implied by previous distance estimates of
50-100~pc. If the MBM12 cloud were at 65~pc as previously reported, its 
associated T~Tauri stars would fall near the main sequence ($\gtrsim100$~Myr).

\item
I have used Li measurements and a combination of H-R diagrams and the
evolutionary models of Baraffe et al.\ (1998) to estimate the absolute and
relative ages of MBM12A and other young associations. These data indicate
ages of $2^{+3}_{-1}$~Myr for MBM12A and 10~Myr for TWA and $\eta$~Cha. 
I also find ages of $30\pm5$ and $30^{+20}_{-5}$~Myr for IC~2602 and HorA,
which are consistent with previous estimates.

\item 
Based on published $JHKL$ photometry, the frequencies and sizes of $L$-band 
excess emission are similar among MBM12A, IC~348, and Taurus. 
Meanwhile, stars in Taurus exhibit greater $K$-band excess emission than
those in the other two regions, suggesting that the inner disks are more 
evolved in the latter. 

\item
The reddenings, radial velocities, and locations of the young stars 
relative to one another and to those of the MBM12 cloud 
combined with the coevality of the stars on the H-R diagram 
are compelling evidence that these stars comprise
an association and that it originated from the cloud.
Most of the young stars in Taurus are located in loose aggregates of 
10-20 young stars within the dark clouds (G\'{o}mez et al.\ 1993).
Similarly, the MBM12A stars form a small, sparse group around the MBM12 cloud.
However, compared to the Taurus sites, the MBM12 cloud is less opaque
($A_V<5$-10 vs.\ $A_V<20$) and the stars are less reddened 
($A_V=0$-2 vs.\ $A_V=0$-10), have more evolved disks and appear to be slightly 
older on the H-R diagram (2~Myr vs.\ 1~Myr), and have a lower space 
density ($n=0.1$~pc$^{-3}$ vs.\ $n=1$-10~pc$^{-3}$).
In addition, the MBM12 cloud may be gravitationally unbound
and in the process of dispersing following the episode 
of star formation that created its associated young stars.
The MBM12 region appears to be a slightly evolved version
of the Taurus star-forming clouds ($\sim1$~Myr) near the age of the IC~348 
cluster ($\sim2$~Myr).

\end{enumerate}

\acknowledgements
Discussions with Lee Hartmann, Scott Wolk, Ray Jayawardhana, Charles Lada,
George Rieke, and Thomas Hearty are appreciated. 
I thank George Rieke and Steward Observatory for use of their facilities. 
I am grateful to France Allard, Isabelle Baraffe, and Franca D'Antona for
access to their most recent calculations.  I was supported by a 
postdoctoral fellowship at the Harvard-Smithsonian Center for Astrophysics.  
This publication makes use of data products from the Two Micron All Sky Survey, 
which is a joint project of the University of Massachusetts and the
Infrared Processing and Analysis Center, funded by the National Aeronautics and 
Space Administration and the National Science Foundation.
Some of the data presented herein were obtained at
the W. M. Keck Observatory, which is operated as a scientific partnership
among the California Institute of Technology, the University of California,
and the National Aeronautics and Space Administration.
The Observatory was made possible by the generous financial support of
the W. M. Keck Foundation. Finally,
I wish to extend special thanks to those of Hawaiian ancestry on whose sacred
mountain we are privileged to be guests. Without their generous hospitality,
some of the observations presented herein would not have been possible.

\appendix

\section{Adopted Data for Young Associations}
\label{sec:append}

Described below are the sources of the Li equivalent widths, effective 
temperatures, and bolometric luminosities used for MBM12A and the other 
young associations in \S~\ref{sec:age}. The detailed procedures given 
for MBM12A also apply to the other populations unless otherwise noted.
In general, for each population I have selected the members that have resolved
photometry and Li measurements and have spectral types between G0 and M6.

\subsection{MBM12A}
\label{sec:mbm12}

\subsubsection{Lithium Strengths}

The Li strengths measured in this work and in previous studies for the members
of MBM12A are listed in Table~\ref{tab:lines}. 
No Li data are available for the new members MBM12A~8, 9, and 11.
For reasons discussed in \S~\ref{sec:sptypes}, the adopted Li strengths are 
those obtained at the highest spectral resolution among the available data.

In actively accreting young stars, continuum veiling from blue excess emission
can weaken the observed strengths of optical absorption features such as 
Li~I 6707.8~\AA.
As a result, the observed values of Li equivalent widths must be corrected for
this effect to arrive at true line strengths, where veilings are typically
measured through analysis of multiple photospheric lines in high-resolution
spectra and defined as $r=I({\rm excess})/I({\rm star})$
(e.g., Hartigan, Edwards, \& Ghandour 1995).
Continuum veiling should be significant only for stars showing clear
signatures of active accretion, such as strong H$\alpha$ emission 
or excess emission at
ultra-violet or IR wavelengths. No veiling correction is made for 
MBM12A~1 and RXJ0306.5+1921, which exhibit weak H$\alpha$ emission and no
IR excess emission (Jayawardhana et al.\ 2001). For MBM12A~5, the presence of 
a disk is implied by mid-IR photometry, but the low H$\alpha$ strength 
suggests that any accretion is not currently significant and that optical 
veiling is probably negligible. In the spectrum of MBM12A~4 obtained in this
study, the H$\alpha$ emission is modest and no significant veiling is found
($r\lesssim0.1$). For MBM12A~3, a veiling of $r=0.3\pm0.1$ is measured 
near the Li transition from the moderate-resolution spectrum presented here. 
At spectral types of late M, the broad, deep molecular bands allow for 
rough veiling estimates from low-resolution spectra. In this way, a veiling 
of $R=0.5\pm0.1$ is measured from the spectrum of the M4.5 star MBM12A~6.
This spectrum was obtained at a different epoch than the adopted Li measurement
by Jayawardhana et al.\ (2001). However, because the H$\alpha$ emission
strength is similar between the two observations, the veiling probably 
did not vary significantly either. Thus, the veiling measured here is used
to correct the Li strength of Jayawardhana et al.\ (2001). 
For MBM12A~7 and 10, the moderate- and low-resolution spectra in this 
work imply veilings of $r\sim0$ and $0.2\pm0.1$ and provide the adopted Li data.
For MBM12A~2 and 12, spectra were not obtained here and no veiling 
measurements have been reported previously.  The strong H$\alpha$ and IR excess 
emission for both stars suggests that significant veiling may be present.
The Li measurements of these two stars could be corrected for veilings 
that are inferred from the H$\alpha$ emission strengths (Strom et al. 1989; 
Hartmann \& Kenyon 1990). However, Basri et al.\ (1991) found
a large scatter between veilings implied by H$\alpha$ and those derived 
from high-resolution spectral analysis. Therefore, rather than use this
method, no veiling correction is attempted for MBM12A~2 and 12 and the 
observed Li strengths are taken as lower limits. The adopted Li 
measurements and the values after deveiling are given in Table~\ref{tab:lines}. 

\subsubsection{Extinctions}
\label{sec:ext}

In the following analysis, standard dwarf colors are taken from the 
compilation of Kenyon \& Hartmann (1995) for types earlier than M0 and 
from the young disk populations described by Leggett (1992) for types 
of M0 and later. The IR colors from Kenyon \& Hartmann (1995) are 
transformed from Johnson-Glass to CIT (Bessell \& Brett 1988).
The $JHK_s$ data in Table~\ref{tab:mem} are from the 2MASS survey. 
For the range of colors of the MBM12A stars ($J-H=0.5$-1), 
the photometric system of 2MASS is very similar to CIT (Carpenter 2001). 
Reddenings are calculated with the extinction law of Rieke \& Lebofsky (1985),
where $A_J=0.282$~$A_V$, $A_J=0.175$~$A_V$, and $A_{K_s}=0.116$~$A_V$ 
(extrapolated from $A_K$).
For the filters used here, Luhman (2000) measured $E(I-K_s)/E(J-H)=4.0$. 
The combination of these relations implies $A_I=0.544$~$A_V$.

In the common method for estimating the reddening towards a young star, 
a color excess is calculated by assuming the intrinsic color, typically 
that of a standard dwarf at the spectral type in question. To ensure that
the color excess reflects only the effect of reddening, 
contamination from short or long wavelength excess emission is minimized
by selecting colors between the $R$ and $H$ bands. Because $R-I$ is 
less susceptible to excess emission than $J-H$, the former is used 
for measuring reddening in this study when it is available. 
Luhman (1999) found that the intrinsic $R-I$ colors of young stars at
M4-M5 types are slightly bluer than the dwarf values, which is a departure
from dwarf colors 
towards giant-like colors. The same effect is present for the 5 members of
MBM12A at M4-M6, whose observed colors are equal to or bluer than dwarf
colors. Therefore, although dereddening the colors of these stars 
to dwarf values implies no reddening, they could in fact have small reddenings 
of up to $A_V\sim0.5$ if the intrinsic colors are even bluer and more giant-like
than the observed colors. However, because the 4 stars at M5.5 and M5.75
have nearly identical colors, it's likely that they are unreddened stars
rather than 4 stars all with the same extinction.

Optical colors were not measured for MBM12A~1, 5, and 12. 
The first two stars were saturated in the $R$ data and the latter source was 
outside of the field that was imaged. 
Because MBM12A~1 shows no IR excess in the colors $H-K_s$, $K_s-L$, and 
$K_s-N$, the $J-H$ color should be uncontaminated by IR excess emission. 
Therefore, the extinction for this star is calculated by dereddening the
observed $J-H$ color to the dwarf value at its spectral type. On the other
hand, MBM12A~5 and 12 both exhibit strong excesses at $L$ and $N$ that 
are indicative of circumstellar disks. 
To estimate reddenings for these stars, the $J-H$ and $H-K$ colors are 
dereddened to the locus observed for classical T Tauri stars (CTTS) 
by Meyer et al.\ (1997). The locus is constructed here such
that the origin coincides with the dwarf colors for the star's spectral type
and the slope is that measured by Meyer et al.\ (1997) for M0 stars, which they 
predicted to remain relatively constant with spectral type. This method 
assumes that the central stars of young mid- and late-M stars have dwarf-like 
$J-H$ and $H-K$ colors, which is suggested by the work of Luhman (1999). 
The reddenings derived from IR colors in this manner will have higher 
uncertainties than those estimated from optical measurements. For instance,
the intrinsic CTTS $J-H$ colors can vary by $\pm0.1$~mag, which corresponds to 
$\pm1$~mag in $A_V$. 

\subsubsection{Effective Temperatures and Bolometric Luminosities}
\label{sec:teff}

For reasons described in previous studies (e.g., Luhman 1999), $I$ and $J$
are the preferred bands for measuring bolometric luminosities of young 
low-mass stars. The luminosities in Table~1 are computed from standard 
dwarf bolometric corrections (see Luhman 1999), the dereddened $J$-band 
measurements, and a distance modulus of 7.2 (\S~\ref{sec:distance}). 
Because the reddening of the MBM12A stars is low and is well-determined by
optical colors and because the luminosity is estimated from a 
wavelength range that suffers little extinction, the uncertainties in
the bolometric luminosities from the dereddening procedure are small. 
Combining the uncertainties in the reddenings, photometry, and bolometric
corrections, the typical errors in the relative bolometric luminosities 
are $\pm0.05$ in log~$L_{\rm bol}$. 
The errors in the absolute luminosities are dominated by the uncertainty in
the distance modulus to the association, which corresponds to 
$\pm0.2$ in $L_{\rm bol}$.

Spectral types of M0 and earlier are converted to effective temperatures with 
the dwarf temperature scale of Schmidt-Kaler (1982).  For spectral types later 
than M0, I adopt the temperature scale developed by Luhman (1999)
for use with the evolutionary models of Baraffe et al.\ (1998).  Detailed 
discussions of the temperature scales and evolutionary models for young
low-mass stars are found in Luhman \& Rieke (1998), Luhman (1999), and 
Luhman et al.\ (2000). 

The reddenings, adopted spectral types, effective temperatures, and bolometric 
luminosities of the members of MBM12A are given in 
Table~\ref{tab:mem}. Photometric measurements from this work, the 2MASS 
survey, and Jayawardhana et al.\ (2001) are also provided. 

\subsection{$\eta$~Cha}

I consider 10 of 12 $\eta$~Cha members from Mamajek et al.\ (1999).
The adopted Li strengths and spectral types for $\eta$~Cha members 
RECX~1, 7, 10, and 12 are from Covino et al.\ (1997).
Data for the remaining $\eta$~Cha stars are from Mamajek et al.\ (1999).
Limited data are available with which to determine the likelihood of blue 
excess emission that could veil the Li feature. Only 2 of the 10 members 
considered here show H$\alpha$ strengths that exceed 10~\AA. At the late M
spectral types of these two stars, H$\alpha$ equivalent widths are larger for
a given line flux because of the depression of surrounding continuum by
molecular bands. Thus, the H$\alpha$ measurements of the $\eta$~Cha members 
suggest that continuum veiling is probably not significant.

Reddenings for the members of $\eta$~Cha are calculated from the $V-I$ 
colors reported by Lawson et al.\ (2000). These colors were used rather 
than $R-I$ data because the latter were not corrected for the effects of
star spots in that study. The resulting extinctions had a range of $A_V=0$-0.8
and an average value of $A_V=0.4$. In contrast, Lawson et al.\ (2000) used 
an assumption of zero reddening and the relations between optical colors 
and spectral types of standard dwarfs to infer spectral types from the colors 
of the $\eta$~Cha members. They claimed that these photometrically derived
spectral types were more accurate than the classifications from spectra 
(Covino et al.\ 1997; Mamajek et al.\ 1999)
because of the relatively high precision of color measurements. However, there
are several problems with estimating spectral types in that manner.  
First, although the colors may be precise, the scatter in the
relations between colors and spectral types are much larger. In other words,
a given measured color -- regardless of its accuracy -- can arise from
a relatively wide range of spectral types. 
Also, the method of Lawson et al.\ (2000) is highly dependent on the assumption
that the relation of colors to spectral types for young stars is the same
as that for dwarfs, which is not the case for certain colors and ranges
of spectral types. For instance, because the optical colors of 
RECX~5 were bluer than the values expected for the spectral classification
of M5, Lawson et al.\ (2000) concluded that the star must be M4. However, 
young stars at these spectral types in fact exhibit intrinsic colors that 
are bluer than those of dwarfs and are more giant-like (Luhman 1999). 
When comparing their photometrically inferred spectral types
to spectral classifications, Lawson et al.\ (2000) found that their 
types were systematically later by an average of one subclass. They adopted
the photometric spectral types and offered no explanation for why the spectral
classifications would have such a systematic error. Instead, it is more
reasonable and likely that the spectral classifications are correct and
that the red color excesses are simply to due to extinction. The small amount of
reddening in question could easily arise from dust within the cluster or
from circumstellar material near these young stars.

Mamajek et al.\ (2000) also note that HD75505 is redder than expected for its 
spectral type of A1 by 0.13~mag in $B-V$, which is equivalent to $A_V=0.4$.
Again, they attribute the color excess to a spectral misclassification
rather than reddening. Mamajek et al.\ (2000) contend that if the spectral
type of this star is truly A1, than it would fall 0.05~dex below the 100~Myr 
isochrone and thus have an age that is inconsistent with that of the 
association. But, of course, at a spectral type near A0, isochrones for
10 and 100~Myr are nearly coincident (e.g., D'Antona \& Mazzitelli 1997). 
Thus, this position on the HR diagram 
is consistent with the age of 10~Myr found for the other members.
Furthermore, Mamajek et al.\ (2000) argue that the red $V-J$ and $V-K$ colors
of HD75505 are evidence of a later spectral type of A6.
However, just as with the $B-V$ colors, these red colors can be explained by
reddening; an A6 star with no reddening and an A0 star with $A_V\sim0.5$ 
will have similar colors in $V-J$ and $V-K$.
It is unlikely that separate studies would spectroscopically misclassify stars 
from B to late M in the amount necessary to mimic reddening of $A_V\sim0.5$.
Therefore, the reddenings computed with the $V-I$ colors are adopted here. 
The bolometric luminosities are derived by combining the dereddened 
$I$ photometry, the appropriate bolometric corrections, and the 
{\it Hipparcos} distance of 97~pc to the star $\eta$~Cha.

\subsection{TWA}

For the H-R diagram, I consider the 14 TWA stars from Webb et al.\ (1999) 
that have resolved spectral types. Because little photometry has been
published for the two members from Sterzik et al.\ (1999), 
they are not included in the H-R diagram. The new members from Zuckerman et
al.\ (2001) are omitted as well as their distances are uncertain and may 
be larger than the previously known members (70-100~pc; Zuckerman et al.\ 2001).
For the Li analysis, all members from these studies that have resolved Li 
measurements and spectral types are included. 

For TWA~1, 2A, 4, and 5A, the adopted Li strengths are from Torres et 
al.\ (2000) and the adopted spectral types are the average of the values from 
Torres et al.\ (2000) and Webb et al.\ (1999). 
The Li data and spectral types for TWA~3A, 3B, 6, 7, 8A, 8B, 9A, 9B, and 10 are
from Webb et al.\ (1999) and for TWA~11B (HR4796B) are from Stauffer et
al.\ (1995). The two members of TWA showing the strongest H$\alpha$ emission
are TWA~1 (TW~Hya) and TWA~3A (Hen~3-600A) with $W_{\lambda}=220$ and 22~\AA, 
respectively. Muzerolle et al.\ (2000) concluded that these stars are probably 
the only known members that have accreting gas disks. Using high-resolution
spectroscopy, they derived a veiling of $r\sim0.2$ at 7000~\AA\ for TW~Hya
and negligible veiling for Hen~3-600A. This veiling for TW~Hya is used to
correct the Li measurement by Torres et al.\ (2000).

Colors from $U$ through $I$ have been measured by Torres et al.\ (2000) 
for 6 stars in TWA. Reddenings have been derived from the $R-I$ colors of
5 of these 6 stars.
For TWA~1 (TW~Hya), the reddening is estimated from the $V-I$ color as the 
$R$ band is contaminated by very strong H$\alpha$ emission. 
The resulting values are $A_V\sim1.2$ for 4 stars and 
$A_V\sim0.2$ for the other 2 sources. 
As with $\eta$~Cha, although the members of TWA are usually assumed to have 
no reddening in previous studies (Webb et al.\ 1999), some do in fact 
have colors indicative of extinction, as noted recently by Zuckerman et 
al.\ (2001). The presence of reddening in these TWA
stars is supported by the fact that no reddening was found in the photometry 
for HorA, which was measured in the same study as the TWA data. 
The TWA stars that lack optical colors must be assumed to have no reddening. 
{\it Hipparcos} distances are available for four TWA members. 
For the remaining members, a distance of 50~pc was adopted. 
The luminosities are derived by combining the distances, dereddened 
$J$ photometry from Webb et al.\ (1999), and bolometric corrections.

\subsection{HorA}

For HorA, I consider the 10 ``probable" members from Torres et al.\ (2000) and
take the Li data, spectral types, and photometry from that study. 
The colors from $B$ through $I$ are consistent with no reddening ($A_V<0.2$)
for the members of HorA. The trigonometric and kinematic distances reported
by Torres et al.\ (2000) are adopted, which have an average value of 60~pc. 
The luminosities are calculated by combining these distances, bolometric 
corrections, and the $I$ photometry. 

\subsection{IC~2602}

\begin{figure}
\plotone{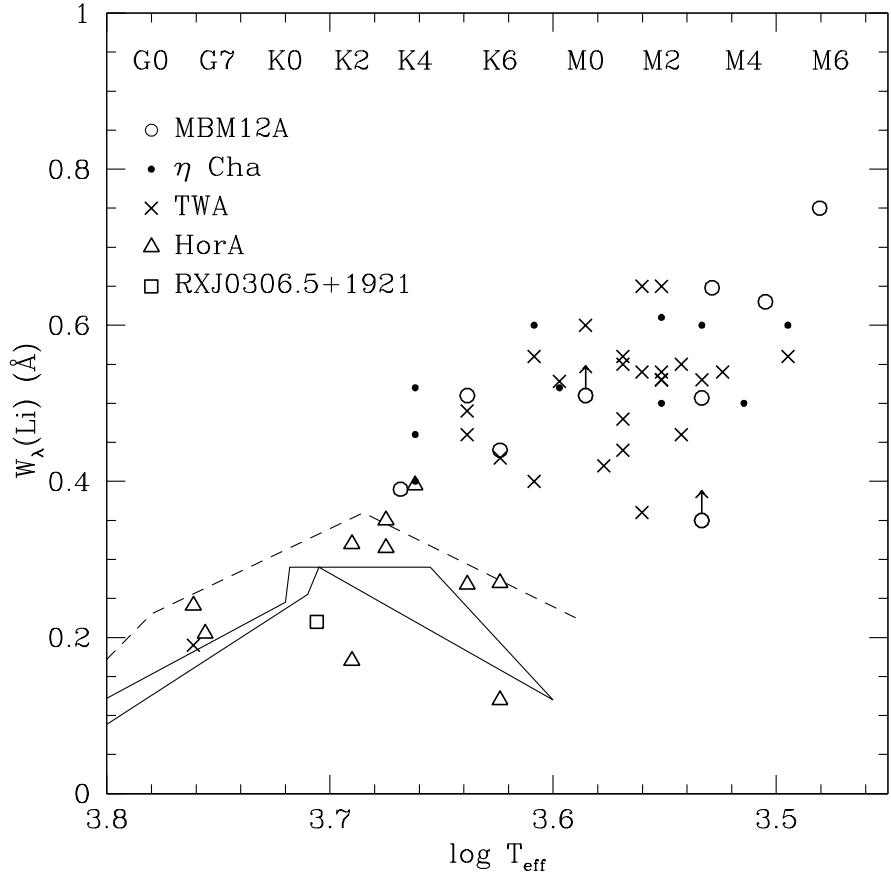}
\caption{
Log~$T_{\rm eff}$ vs.\ $W_{\lambda}$(Li) for the young associations 
MBM12A, $\eta$~Cha, TWA, and HorA.
Li measurements are not available for MBM12A~8, 9, and 11.
RXJ0306.5+1921 is $2\arcdeg$ east of the MBM12 cloud and has been suggested 
as a member, but it is probably an older, unrelated star. 
Shown for comparison are the upper envelopes for rapidly and slowly rotating
stars in the Pleiades (125~Myr; {\it upper and lower solid lines}) and stars
in IC~2602 (30 Myr; {\it dashed line}) (see Neuh\"{a}user et al.\ 1997).
}
\label{fig:li}
\end{figure}

\begin{figure}
\epsscale{0.75}
\plotone{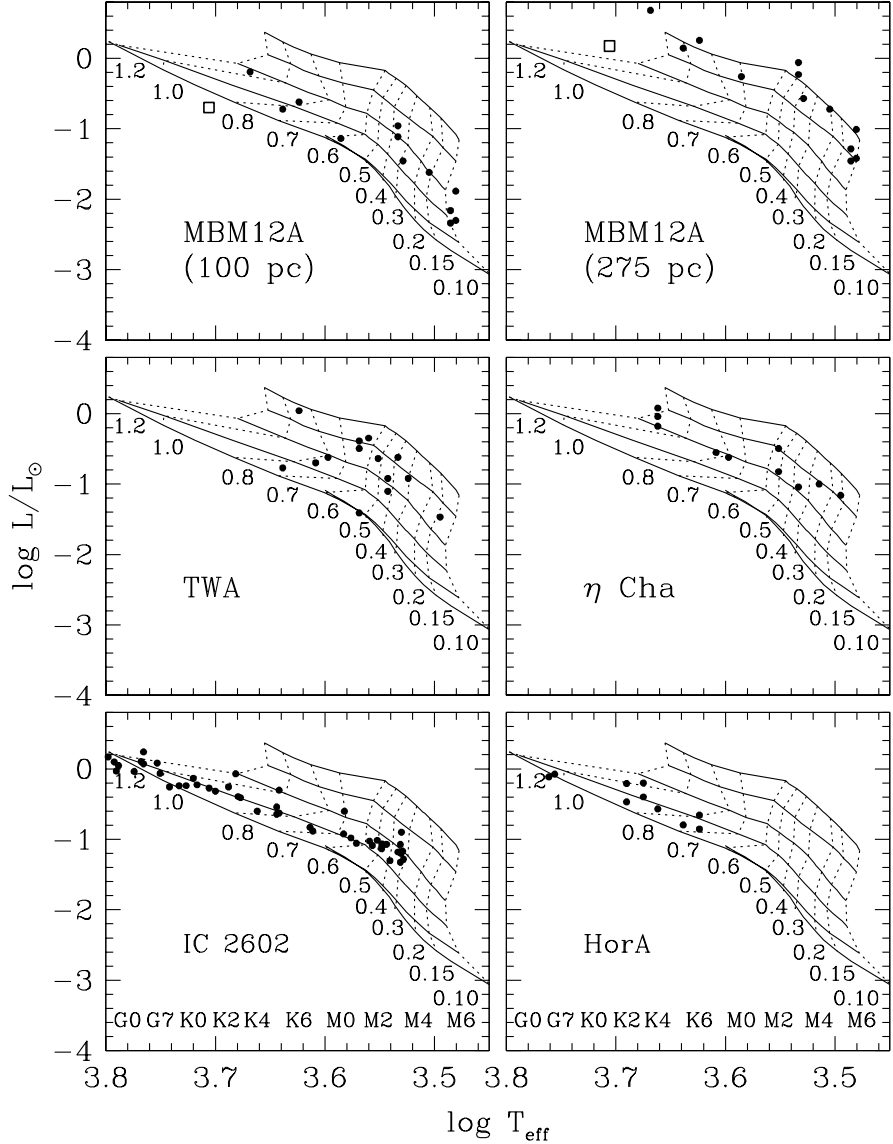}
\caption{
H-R diagrams for the young populations MBM12A, $\eta$~Cha, TWA, HorA, and 
IC~2602. The MBM12A members are plotted for the distance of 275~pc derived
here and for the previously published upper limit of 100~pc.
RXJ0306.5+1921 ({\it open square}) is $2\arcdeg$ east of the MBM12 cloud 
and has been suggested as a member, but it is probably an older, unrelated star.
The theoretical evolutionary models of Baraffe et al.\ (1998) are shown, where 
the horizontal solid lines are isochrones representing ages of 1, 3, 10, 30, 
and 100~Myr and the main sequence, from top to bottom.
The M spectral types have been converted to effective temperatures with a 
scale that is compatible with these evolutionary models at young ages
(Luhman 1999), which is intermediate between the scales for M dwarfs and giants.
}
\label{fig:hr}
\end{figure}

\begin{figure}
\plotone{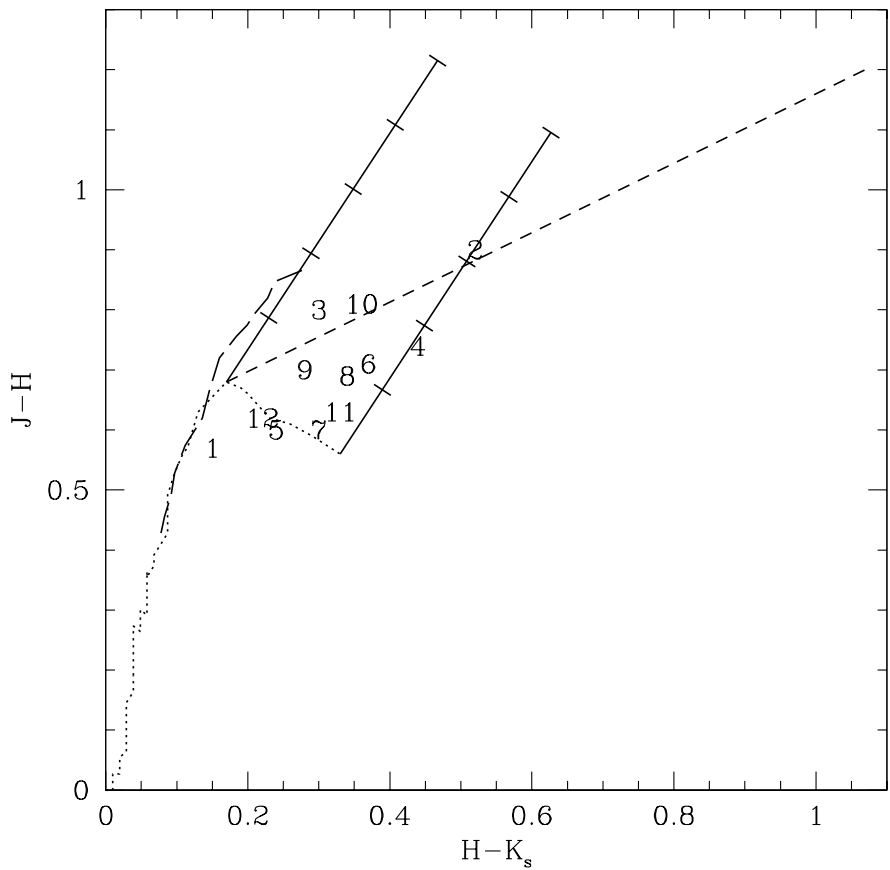}
\caption{
$H-K_s$ vs.\ $J-H$ for young stars near the MBM12 cloud.
MBM12A~1-6 are previously known young stars and MBM12A~7-11 are the new
ones discovered in this work. 
The young star MBM12A~12 (S18) is located $2.5\arcdeg$ southeast of 
the MBM12 cloud and is a likely member of this association.
The colors have been corrected for the reddenings in Table~\ref{tab:mem}. 
For most of the stars, the extinctions are computed from $R-I$ colors.
The exceptions are MBM12A~1, 5, and 12, for which the intrinsic $JHK_s$ 
colors expected 
for main sequence stars or CTTS were assumed to derive extinctions (see text).
The main sequence ({\it dotted line}), giant sequence ({\it long dashed line}), 
and locus of M0 CTTS in Taurus (Meyer et al.\ 1997; {\it short dashed line}) 
are plotted with reddening vectors ({\it solid lines}) originating at M0V 
and M6V, which are marked at intervals of $A_V=1$.
}
\label{fig:jhhk}
\end{figure}

\begin{figure}
\plotone{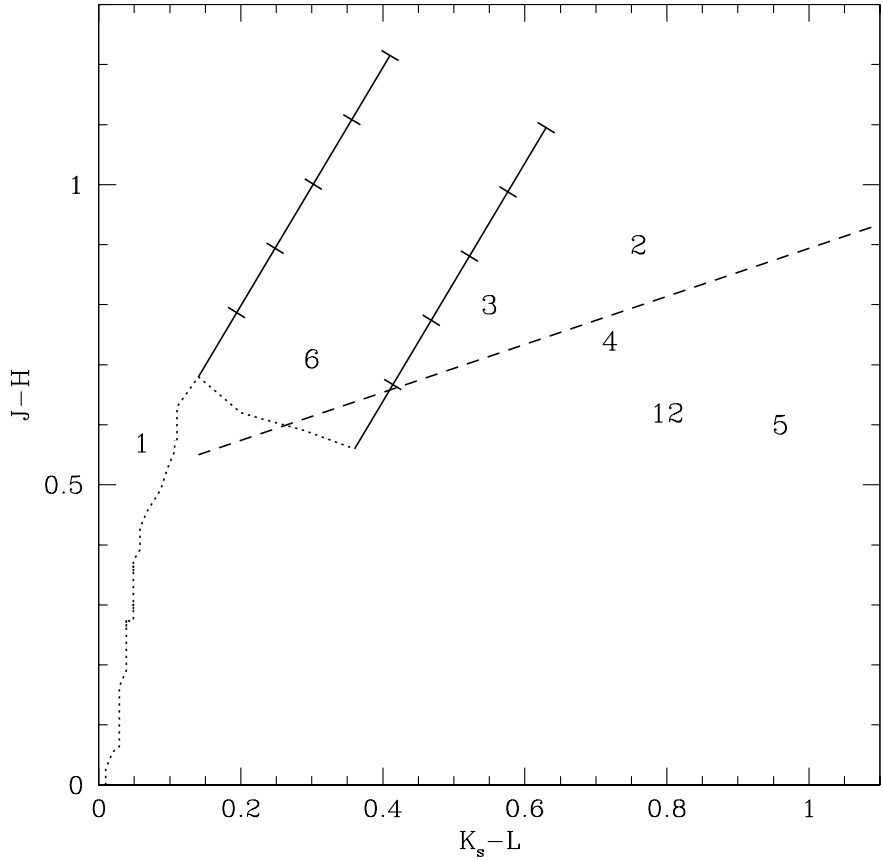}
\caption{
$K_s-L$ vs.\ $J-H$ for young stars near the MBM12 cloud.
MBM12A~1-6 and 12 are the only members of the association for which $L$ 
data are available (Jayawardhana et al.\ 2001). 
The young star MBM12A~12 (S18) is located $2.5\arcdeg$ southeast of 
the MBM12 cloud and is a likely member of this association.
The colors have been corrected for the reddenings in Table~\ref{tab:mem}. 
The main sequence ({\it dotted line})
and locus of M0 CTTS in Taurus (Meyer et al.\ 1997; {\it short dashed line}) 
are plotted with reddening vectors ({\it solid lines}) originating at M0V 
and M6V, which are marked at intervals of $A_V=1$.
}
\label{fig:jhkl}
\end{figure}

In Figure~\ref{fig:li}, the upper envelopes for Li strengths in rapidly and 
slowly rotating stars in the Pleiades (125~Myr; Stauffer, Schultz \& Kirkpatrick
1998) and in stars in IC~2602 (30 Myr; Stauffer et al.\ 1997) have been defined 
by Neuh\"{a}user et al.\ (1997) for Li measurements from Soderblom et 
al.\ (1993), Garc\'{\i}a L\'{o}pez, Rebolo, \& Mart\'{\i}n (1994), Randich et 
al.\ (1997), and Stauffer et al.\ (1997). There is no evidence of disk 
accretion or continuum veiling in these relatively old stars ($\geq30$~Myr).

Temperatures and luminosities for the likely members of IC~2602 (Stauffer 
et al.\ 1997; Randich et al.\ 2001) were derived by combining the relations 
between spectral types, effective temperatures, colors, and bolometric 
corrections in \S~\ref{sec:mbm12}, $V$ photometry and $V-I_C$ colors, a 
distance modulus of 5.95, and a reddening of $E(B-V)=0.04$ (Stauffer et 
al.\ 1997).

\newpage

\clearpage
 
\clearpage
 
\clearpage
 
\clearpage
 
\clearpage
 
\clearpage
 
\clearpage
 
\clearpage
 
\clearpage

\clearpage
 
\clearpage
 
\clearpage

\clearpage
 
\clearpage
 
\begin{deluxetable}{llllllllllllllll}
\tabletypesize{\scriptsize}
\rotate
\tablewidth{0pt}
\tablecaption{Members of the MBM12 Association \label{tab:mem}}
\tablehead{\colhead{MBM12A}
& \colhead{Other Names}
& \colhead{$\alpha$(2000)} & \colhead{$\delta$(2000)}
& \colhead{Spectral Type\tablenotemark{a}} 
& \colhead{Adopt} & \colhead{$T_{\rm eff}$\tablenotemark{b}}
& \colhead{$A_J$} & \colhead{$L_{\rm bol}$} & \colhead{$R-I$} & \colhead{$I$}
& \colhead{$J-H$} & \colhead{$H-K_s$} & \colhead{$K_s$} & \colhead{$K_s-L$} & \colhead{$K_s-N$}\\
}
\startdata
1 & RXJ0255.4+2005 & 2\ 55\ 25.78 & 20\ 04\ 51.7 & K6(2) & K6 &  4205 &  0.11 & 1.8 & \nodata & 10.69 & 0.62 & 0.18 &  8.99 & 0.08 & 0.15 \\
2 & LkHa262 & 2\ 56\ 07.99 & 20\ 03\ 24.3 & M0(1,2) & M0 &  3850 &  0.33 & 0.55 & 1.00 & 12.51 & 1.03 & 0.60 &  9.66 & 0.82 & 2.64 \\
3 & LkHa263 & 2\ 56\ 08.42 & 20\ 03\ 38.6 & M2(1),M4(2),M3(4) & M3 &  3415 &  0.00 & 0.59 & 1.32 & 12.17 & 0.80 & 0.30 &  9.52 & 0.55 & 2.45 \\
4 & LkHa264 & 2\ 56\ 37.56 & 20\ 05\ 37.1 & K5(1,2),K5-K6(4) & K5 &  4350 &  0.24 & 1.4 & 0.81 & 11.60 & 0.83 & 0.49 &  8.90 & 0.77 & 3.75 \\
5 & E02553+2018 & 2\ 58\ 11.23 & 20\ 30\ 03.5 & K3(3),K4(2) & K3.5 &  4660 &  0.55 & 4.8 & \nodata & 10.54 & 0.80 & 0.36 &  8.11 & 1.06 & 2.45 \\
6 & RXJ0258.3+1947 & 2\ 58\ 16.09 & 19\ 47\ 19.6 & M5(2),M4.5(4) & M4.5 &  3198 &  0.00 & 0.19 & 1.30 & 13.40 & 0.71 & 0.37 & 10.73 & 0.30 & 2.15 \\
7 & RXJ0256.3+2005 & 2\ 56\ 17.98 & 20\ 06\ 09.9 & M5.75(4) & M5.75 &  3024 &  0.00 & 0.098 & 1.89 & 14.70 & 0.60 & 0.30 & 11.66 & \nodata & \nodata \\
8 & \nodata & 2\ 57\ 49.02 & 20\ 36\ 07.8 & M5.5(4) & M5.5 &  3058 &  0.00 & 0.035 & 1.90 & 15.85 & 0.69 & 0.34 & 12.61 & \nodata & \nodata \\
9 & \nodata & 2\ 58\ 13.37 & 20\ 08\ 25.0 & M5.75(4) & M5.75 &  3024 &  0.00 & 0.038 & 1.91 & 15.77 & 0.70 & 0.28 & 12.60 & \nodata & \nodata \\
10 & \nodata & 2\ 58\ 21.10 & 20\ 32\ 52.7 & M3.25(4) & M3.25 &  3379 &  0.05 & 0.27 & 1.45 & 13.04 & 0.83 & 0.37 & 10.33 & \nodata & \nodata \\
11 & \nodata & 2\ 58\ 43.80 & 19\ 40\ 38.3 & M5.5(4) & M5.5 &  3058 &  0.00 & 0.052 & 1.93 & 15.30 & 0.63 & 0.33 & 12.25 & \nodata & \nodata \\
12 & S18 & 3\ 02\ 21.05 & 17\ 10\ 34.2 & M3(2) & M3 &  3415 &  0.50 & 0.87 & \nodata & \nodata & 0.81 & 0.33 &  9.56 & 0.90 & 1.75 \\
\enddata
\tablenotetext{a}{
1=Herbig \& Bell 1988, 2=Hearty et al.\ 2000b, 3=Fleming, Gioia, \& Maccacaro 
1989, 4=this work.}
\tablenotetext{b}{Intermediate temperature scale of Luhman 1999.}
\tablecomments{ $RI$, $JHK_s$, and $LN$ data are from this work, the 2MASS 
survey, and Jayawardhana et al.\ 2001, respectively. 
Coordinates for MBM12A~1-11 are measured from the $I$-band images, 
where the plate solutions were derived using 2MASS coordinates of sources 
appearing in the images. Coordinates for MBM12A~12 are directly from the
2MASS survey. 
}
\end{deluxetable}

\clearpage

\begin{deluxetable}{llllllll}
\tabletypesize{\footnotesize}
\tablewidth{0pt}
\tablecaption{Background Stars \label{tab:back}}
\tablehead{
\colhead{ID} & \colhead{$\alpha$(2000)} & \colhead{$\delta$(2000)}
& \colhead{$R-I$} & \colhead{$I$} & \colhead{$J-H$} & \colhead{$H-K_s$} 
& \colhead{$K_s$}}
\startdata
2MASSs J0253511+202808 & 2\ 53\ 51.15 & 20\ 28\ 08.5 & 0.91 & 12.08 & 0.64 & 0.19 & 10.03 \\
2MASSs J0253530+194517 & 2\ 53\ 53.09 & 19\ 45\ 18.0 & 0.94 & 11.25 & 0.78 & 0.22 &  8.91 \\
2MASSs J0254108+201804 & 2\ 54\ 10.90 & 20\ 18\ 04.5 & 0.94 & 11.47 & 0.72 & 0.21 &  9.29 \\
2MASSs J0254151+202218 & 2\ 54\ 15.13 & 20\ 22\ 18.6 & 0.94 & 12.24 & 0.76 & 0.22 &  9.90 \\
2MASSs J0254471+202918 & 2\ 54\ 47.19 & 20\ 29\ 18.5 & 0.84 & 11.18 & 0.62 & 0.21 &  9.16 \\
2MASSs J0255119+201624 & 2\ 55\ 11.98 & 20\ 16\ 24.5 & 1.09 & 11.97 & 0.63 & 0.23 &  9.67 \\
2MASSs J0255230+194145 & 2\ 55\ 23.10 & 19\ 41\ 45.3 & 1.06 & 11.30 & 0.80 & 0.28 &  8.75 \\
2MASSs J0255386+202443 & 2\ 55\ 38.64 & 20\ 24\ 43.9 & 1.05 & 11.96 & 0.79 & 0.26 &  9.56 \\
2MASSs J0255516+201807 & 2\ 55\ 51.65 & 20\ 18\ 07.9 & 1.24 & 13.08 & 0.78 & 0.22 & 10.64 \\
2MASSs J0256007+194046 & 2\ 56\ 00.72 & 19\ 40\ 46.1 & 1.82 & 12.57 & 1.35 & 0.51 &  7.76 \\
2MASSs J0256087+193344 & 2\ 56\ 08.78 & 19\ 33\ 44.9 & 1.52 & 12.35 & 1.09 & 0.36 &  8.43 \\
2MASSs J0256129+202915 & 2\ 56\ 12.94 & 20\ 29\ 15.7 & 1.11 & 12.74 & 0.80 & 0.30 & 10.04 \\
2MASSs J0256462+193351 & 2\ 56\ 46.24 & 19\ 33\ 51.7 & 0.93 & 12.00 & 0.49 & 0.19 & 10.24 \\
2MASSs J0256481+193204 & 2\ 56\ 48.13 & 19\ 32\ 04.5 & 0.91 & 11.48 & 0.56 & 0.18 &  9.60 \\
2MASSs J0256574+195815 & 2\ 56\ 57.46 & 19\ 58\ 15.6 & 0.90 & 12.34 & 0.54 & 0.17 & 10.44 \\
2MASSs J0256598+200215 & 2\ 56\ 59.85 & 20\ 02\ 15.7 & 1.07 & 12.92 & 0.95 & 0.24 & 10.13 \\
2MASSs J0257004+195124 & 2\ 57\ 00.41 & 19\ 51\ 24.5 & 0.89 & 11.37 & 0.67 & 0.21 &  9.30 \\
2MASSs J0257038+201311 & 2\ 57\ 03.82 & 20\ 13\ 11.2 & 0.97 & 11.93 & 0.69 & 0.26 &  9.60 \\
2MASSs J0257159+202616 & 2\ 57\ 15.94 & 20\ 26\ 16.7 & 0.88 & 12.28 & 0.60 & 0.24 & 10.18 \\
2MASSs J0257214+194730 & 2\ 57\ 21.49 & 19\ 47\ 30.8 & \nodata & \nodata & 0.66 & 0.13 &  8.80 \\
2MASSs J0257280+195738 & 2\ 57\ 28.08 & 19\ 57\ 38.3 & 1.05 & 11.26 & 0.79 & 0.21 &  8.68 \\
2MASSs J0257299+204308 & 2\ 57\ 29.94 & 20\ 43\ 08.9 & 1.88 & 18.20 & 1.16 & 0.51 & 13.64 \\
2MASSs J0257391+193107 & 2\ 57\ 39.11 & 19\ 31\ 07.3 & 1.32 & 12.92 & 0.87 & 0.35 &  9.85 \\
2MASSs J0257439+193336 & 2\ 57\ 43.90 & 19\ 33\ 36.6 & 0.82 & 11.63 & 0.66 & 0.18 &  9.71 \\
2MASSs J0257460+200055 & 2\ 57\ 46.05 & 20\ 00\ 55.7 & 1.22 & 13.13 & 0.64 & 0.28 & 10.56 \\
2MASSs J0257474+192609 & 2\ 57\ 47.43 & 19\ 26\ 09.3 & 1.42 & 13.44 & 1.04 & 0.36 & 10.01 \\
2MASSs J0257536+194325 & 2\ 57\ 53.60 & 19\ 43\ 25.2 & 1.49 & 13.03 & 1.11 & 0.44 &  9.22 \\
2MASSs J0257540+202724 & 2\ 57\ 54.09 & 20\ 27\ 24.7 & 0.99 & 12.74 & 0.73 & 0.19 & 10.40 \\
2MASSs J0257562+193228 & 2\ 57\ 56.28 & 19\ 32\ 28.8 & 1.35 & 11.19 & 1.05 & 0.33 &  8.03 \\
2MASSs J0257597+200343 & 2\ 57\ 59.72 & 20\ 03\ 43.5 & 1.16 & 11.71 & 0.78 & 0.26 &  9.02 \\
2MASSs J0258046+193146 & 2\ 58\ 04.69 & 19\ 31\ 46.9 & 1.30 & 12.41 & 0.90 & 0.34 &  9.33 \\
2MASSs J0258157+194626 & 2\ 58\ 15.72 & 19\ 46\ 26.4 & 0.86 & 11.84 & 0.60 & 0.27 &  9.72 \\
2MASSs J0258257+200824 & 2\ 58\ 25.78 & 20\ 08\ 24.9 & 1.19 & 12.54 & 0.78 & 0.22 & 10.09 \\
2MASSs J0258458+201114 & 2\ 58\ 45.81 & 20\ 11\ 15.0 & 1.06 & 13.04 & 0.86 & 0.23 & 10.60 \\
2MASSs J0259057+202309 & 2\ 59\ 05.71 & 20\ 23\ 09.5 & 1.10 & 12.52 & 0.59 & 0.24 & 10.16 \\
2MASSs J0259134+203452 & 2\ 59\ 13.43 & 20\ 34\ 52.6 & 0.76 & 10.84 & 0.42 & 0.17 &  9.18 \\
2MASSs J0259258+201702 & 2\ 59\ 25.83 & 20\ 17\ 03.0 & 0.93 & 12.60 & 0.74 & 0.21 & 10.38 \\
\enddata
\tablecomments{Spectroscopy indicates early or giant spectral types for
all sources except 2MASSs~J0257299+204308, which is classified as M2.75V.}
\end{deluxetable}

\clearpage

\begin{deluxetable}{lllllllll}
\tabletypesize{\scriptsize}
\tablewidth{0pt}
\tablecaption{Foreground Stars \label{tab:fore}}
\tablehead{
\colhead{ID} & \colhead{$\alpha$(2000)} & \colhead{$\delta$(2000)}
& \colhead{Spectral Type}
& \colhead{$R-I$} & \colhead{$I$} & \colhead{$J-H$} & \colhead{$H-K_s$} 
& \colhead{$K_s$}}
\startdata
2MASSs J0252368+193424 & 2\ 52\ 36.80 & 19\ 34\ 24.2 & M0.5V-M1V & 1.09 & 11.88 & 0.63 & 0.20 & 10.01 \\
2MASSs J0252486+203908 & 2\ 52\ 48.63 & 20\ 39\ 08.6 & M2.5V & 1.22 & 13.46 & 0.60 & 0.24 & 11.29 \\
2MASSs J0253030+194904 & 2\ 53\ 03.00 & 19\ 49\ 04.3 & M8.25V & 2.36 & 17.14 & 0.74 & 0.42 & 13.12 \\
2MASSs J0253031+192907 & 2\ 53\ 03.11 & 19\ 29\ 07.6 & M4.75V & 1.85 & 13.25 & 0.61 & 0.26 & 10.68 \\
2MASSs J0253099+203849 & 2\ 53\ 09.92 & 20\ 38\ 49.2 & M0V & 0.97 & 12.58 & 0.67 & 0.18 & 10.57 \\
2MASSs J0253243+200659 & 2\ 53\ 24.37 & 20\ 06\ 59.5 & M3V & 1.42 & 13.90 & 0.60 & 0.30 & 11.61 \\
2MASSs J0254038+193136 & 2\ 54\ 03.87 & 19\ 31\ 36.2 & M3V & 1.43 & 13.75 & 0.59 & 0.27 & 11.63 \\
2MASSs J0255083+193859 & 2\ 55\ 08.31 & 19\ 38\ 59.2 & M5.75V & 2.15 & 17.44 & 0.53 & 0.46 & 14.26 \\
2MASSs J0255214+194614 & 2\ 55\ 21.52 & 19\ 46\ 15.1 & M0.5V-M1V & 0.99 & 12.31 & 0.66 & 0.20 & 10.48 \\
2MASSs J0255350+200648 & 2\ 55\ 35.02 & 20\ 06\ 48.7 & M4.5V & 1.70 & 13.46 & 0.64 & 0.26 & 10.91 \\
2MASSs J0257501+195830 & 2\ 57\ 50.17 & 19\ 58\ 30.1 & M4.5V & 1.74 & 14.75 & 0.66 & 0.25 & 12.05 \\
2MASSs J0257585+202208 & 2\ 57\ 58.56 & 20\ 22\ 08.0 & M4V-M4.5V & 1.58 & 13.51 & 0.61 & 0.29 & 11.00 \\
2MASSs J0258110+192908 & 2\ 58\ 11.03 & 19\ 29\ 08.0 & M9V & 2.51 & 18.84 & 0.55 & 0.54 & 14.91 \\
2MASSs J0258146+202249 & 2\ 58\ 14.65 & 20\ 22\ 49.0 & M3V-M3.5V & 1.45 & 13.16 & 0.60 & 0.23 & 10.89 \\
2MASSs J0258228+200722 & 2\ 58\ 22.88 & 20\ 07\ 22.4 & K7V-K8V & 0.94 & 12.71 & 0.68 & 0.19 & 10.84 \\
2MASSs J0258343+200023 & 2\ 58\ 34.32 & 20\ 00\ 23.6 & M2V & 1.15 & 13.23 & 0.70 & 0.27 & 11.03 \\
2MASSs J0259129+200549 & 2\ 59\ 12.98 & 20\ 05\ 49.3 & M2.5-M3V & 1.34 & 13.13 & 0.55 & 0.26 & 11.08 \\
\enddata
\end{deluxetable}

\clearpage

\begin{deluxetable}{llllllll}
\tablewidth{0pt}
\tablecaption{Candidate Members of MBM12A \label{tab:cand}}
\tablehead{
\colhead{ID} & \colhead{$\alpha$(2000)} & \colhead{$\delta$(2000)}
& \colhead{$R-I$} & \colhead{$I$} & \colhead{$J-H$} & \colhead{$H-K_s$} 
& \colhead{$K_s$}}
\startdata
2MASSs J0252490+195252 & 2\ 52\ 49.09 & 19\ 52\ 52.4 & 1.57 & 14.37 & 0.58 & 0.27 & 12.01 \\
2MASSs J0255056+194454 & 2\ 55\ 05.61 & 19\ 44\ 53.3 & 2.61 & 19.60 & 0.81 & 0.31 & 15.65 \\
2MASSs J0255372+204311 & 2\ 55\ 37.26 & 20\ 43\ 11.2 & 2.56 & 19.60 & 0.77 & 0.37 & 15.74 \\
2MASSs J0255452+192526 & 2\ 55\ 45.24 & 19\ 25\ 26.1 & 2.35 & 19.89 & 1.07 & 0.71 & 15.17 \\
2MASSs J0258344+194650 & 2\ 58\ 34.52 & 19\ 46\ 50.5 & 1.78 & 15.04 & 0.44 & 0.36 & 12.63 \\
2MASSs J0258523+195840 & 2\ 58\ 52.39 & 19\ 58\ 40.0 & 1.79 & 15.29 & 0.66 & 0.30 & 12.43 \\
\enddata
\end{deluxetable}

\clearpage

\begin{deluxetable}{lllllllll}
\tabletypesize{\scriptsize}
\tablewidth{0pt}
\tablecaption{Line Measurements for Members of MBM12A \label{tab:lines}}
\tablehead{\colhead{MBM12A}
& \colhead{$W_{\lambda}$(H$\alpha$)} & \colhead{$W_{\lambda}$(Li)}
& \colhead{$W_{\lambda}$(H$\alpha$)} & \colhead{$W_{\lambda}$(Li)}
& \colhead{$W_{\lambda}$(H$\alpha$)} & \colhead{$W_{\lambda}$(Li)}
& \multicolumn{2}{c}{$W_{\lambda}$(Li)} \\
\colhead{}
& \multicolumn{2}{c}{H00b}
& \multicolumn{2}{c}{Jayawardhana et al.\ 2001}
& \multicolumn{2}{c}{this work}
& \colhead{adopted} & \colhead{deveiled}
}
\startdata
1 & $-1.26$ & $0.44\pm0.01$\tablenotemark{b} & $-0.93$ & $0.31\pm0.03$ & \nodata & \nodata & 0.44 & 0.44 \\
2 & $-32.1$ & $0.29\pm0.09$ & $-39.7$\tablenotemark{a} & $0.51\pm0.07$ & \nodata & \nodata & 0.51 & $\geq0.51$ \\
3 & $-32.9$ & $0.38\pm0.09$ & $-13.3$\tablenotemark{a} & $0.39\pm0.05$ & $-25\pm0.5$\tablenotemark{a} & $0.43\pm0.05$ & 0.39 & 0.51 \\
4 & $-58.9$ & $0.51\pm0.02$\tablenotemark{b} & $-17.7$\tablenotemark{a} & $0.41\pm0.01$ & $-17.5\pm0.5$ & $0.50\pm0.05$ & 0.51 & 0.51 \\
5 & $-1.6$ & $0.62\pm0.09$ & $-3.1$\tablenotemark{a} & $0.39\pm0.01$ & \nodata & \nodata & 0.39 & 0.39 \\
6 & $-24.5$ & $0.58\pm0.09$ & $-33.8$ & $0.42\pm0.04$ & $-29\pm1$ & \nodata & 0.42 & 0.63 \\
7 & \nodata & \nodata & \nodata & \nodata & $-13.5\pm0.5$ & $0.75\pm0.05$ & 0.75 & 0.75 \\
8 & \nodata & \nodata & \nodata & \nodata & $-120\pm20$ & \nodata & \nodata & \nodata \\
9 & \nodata & \nodata & \nodata & \nodata & $-9.5\pm2$ & \nodata & \nodata & \nodata \\
10 & \nodata & \nodata & \nodata & \nodata & $-12\pm0.5$ & $0.54\pm0.05$ & 0.54 & 0.65 \\
11 & \nodata & \nodata & \nodata & \nodata & $-13.5\pm0.5$ & \nodata & \nodata & \nodata \\
12 & $-79.0$ & $0.31\pm0.09$ & $-69.2$ & $0.35\pm0.01$ & \nodata & \nodata & 0.35 & $\geq0.35$ \\
\enddata
\tablenotetext{a}{Double-peaked profile.}
\tablenotetext{b}{High resolution.}
\tablecomments{Positive and negative equivalent widths denote absorption and
emission, respectively, in units of \AA.}
\end{deluxetable}

\end{document}